\documentclass[preprintnumbers,article,amsmath,amssymb,floatfix,10pt,prd,onecolumn,
superscriptaddress,nofootinbib]{revtex4}
\usepackage{bbm}
\usepackage{amsfonts}
\usepackage{mathrsfs}
\usepackage{latexsym}

\usepackage{epsfig}
\usepackage{epstopdf}
\usepackage{graphicx}
\usepackage{amssymb}
\usepackage{amsmath}
\usepackage{dcolumn}
\usepackage{bm}
\usepackage{color}
\usepackage{comment}
\usepackage{xcolor}
\begin{document}

\title{Wormhole Structures in Logarithmic-Corrected \textbf{$R^2$} Gravity }

\author{I. Fayyaz}
\email{iffat845@gmail.com}\affiliation{National University of Computer and
Emerging Sciences,\\ Lahore Campus, Pakistan.}
\author{M. Farasat Shamir}
\email{farasat.shamir@nu.edu.pk}\affiliation{National University of Computer and
Emerging Sciences,\\ Lahore Campus, Pakistan.}

\begin{abstract}
This paper is devoted to find the feasible shape functions for the construction of static wormhole geometry in the frame work of logarithmic-corrected
\textbf{$R^2$} gravity model. We discuss the asymptotically flat wormhole solutions sustained by the matter sources with anisotropic pressure, isotropic pressure and barotropic pressure. For anisotropic case, we consider three shape functions and evaluate the null energy conditions and weak energy conditions graphically along with their regions. Moreover, for barotropic and isotropic pressures, we find shape function analytically and discuss its properties. For the formation of traversable wormhole geometries, we cautiously choose the values of parameters involved in $f(R)$ gravity model.
We show explicitly that our wormhole solutions violates the non-existence theorem even with logarithmic corrections. We discuss all physical properties via graphical analysis and it is concluded that the wormhole
solutions with relativistic formalism can be well justified with logarithmic corrections.

\end{abstract}

\maketitle

\date{\today}

\section{Introduction}


In recent era, the investigation about the existence of wormholes  is an interesting and attractive topic for researchers. Flamm \cite{fla}  proposed the possibility
of such solutions in 1916 but after some time it was realized that his proposed solutions were not relativistically applicable. A detailed investigation about the
possible solutions is given by Einstein and Rosen \cite{ein} in 1935 and introduced a particle as a bridge (also known as Einstein-Rosen Bridge)
connecting asymptotic regions of  a single sheet or two identical sheets. In fact, wormhole is an imaginary topology which is featured as a tube-like tunnel
between different space times away from each other. Theory of general relativity (GR) predicts the presence of exotic matter which seems to formulate the structure of wormholes. Exotic matter is such type of
matter which violates the null energy condition (NEC) at least near the throat of the wormhole. In literature, a lot of articles are available to understand the
physics of wormholes in detail \cite{kim,haw1,hoc1,hoc2}. Wormholes can be characterized as static or non-static
depending on whether the throat is a constant or a variable.

In recent years, accelerated expansion of the universe has become an interesting and thought-provoking topic for researchers \cite{spe,per}. According to the
researchers, GR is a ahead of the game in cosmology but it explains dark energy and acceleration of the
universe in weak regime. Qadir et al. \cite{qad} suggested that the modification of GR may serve the scientific explanation of accelerated expansion of the universe,
flatness issues and the dark energy problems in a better way. Recently, Capozziello et al. \cite{cap1} and Demianski et al. \cite{dim} demonstrated that
these modified theories can provide such models which have the proficiency to reflect the Hubble diagram derived from SNelve surveys. Thus, these modified theories have yield multiple directions of cosmology for researchers.
A well-known theory of modified gravity i.e. $f(R)$ gravity is an extension of Einstein
theory of GR. Buchdahl \cite{buc} proposed $f(R)$ theory of gravity by varying the Einstein Hilbert action with an arbitrary function $f(R)$. Further, Martin et al. \cite{mar} highlighted the issues associated with current cosmic acceleration. Nojiri and Odintsov \cite{noj1}, presented some $f(R)$ models which included higher order
curvature invariants as function of Ricci scalar $R$. Harko et al. \cite{har1} proposed a generalization of $f(R)$ modified theories of gravity with an arbitrary
coupling between matter and geometry. Rahaman and collaborators \cite{rah} explored exact solutions for noncommutative wormholes in $f(R)$ gravity in which matter possessed with Lorentzian Distribution. Harko et al. \cite{har2} shown some results for wormholes in $f(R)$ modification. Pavlovic and Sossich \cite{pav} discussed wormholes for different $f(R)$ models and found solutions which do not demand exotic matter. Bahamonde et al. \cite{bah}  used the approximation of small wormholes and formulated a non-static wormhole asymptotically approaching towards the  Friedmann-Lemaître-Robertson-Walker universe. Zubair et al. \cite{zub1} used three different types of equation of state to discuss wormhole solutions in $f(R, \phi)$ gravity.

Bronnikov and Starobinsky \cite{5bron} discussed that in general no wormhole can be formed if scalar function $f(\phi)$ is positive every where in scalar-tensor models of dark energy (commonly known as non-existence theorem). Further, Bronnikov et al. \cite{bro} demonstrated the wormhole existence in the context of scalar-tensor and $f(R)$ gravity. It was concluded that non-existence theorem could be violated when $f(\phi)$ or equivalently $\frac{df}{dR}=F(R)$ is negative.

Morris and Thorne \cite{mor} were the first who used principles of GR to discuss the static spherically symmetric wormholes.
The space-time which describes the geometry of a static wormhole is known as  Morris-Thorne metric given by:
\begin{equation}\label{4}
ds^2=e^{\psi(r)} dt^2 -e^{\lambda(r)} dr^2 -r^2 d\theta^2 -r^2\sin^2\theta d\phi^2.
\end{equation}
The metric coefficient $\psi(r)$ represents red-shift function which amplify the gravitational red-shift with respect to the radial coordinator $r$ and
\begin{equation}\label{5}
\lambda(r)=-ln [1-\frac{\epsilon(r)}{r}],
\end{equation}
where $\epsilon(r)$ is known as shape function. To construct the wormholes, some key properties related
to metric potential are listed below\\\\
\begin{itemize}
\item{It is assumed that the wormholes are free from event horizon. So, for the existence of traversable wormhole, the red-shift parameter $\psi(r)$ must be
 finite everywhere. }
\item{For the required shape of wormhole throat, it is needed that the radial coordinate must not show the monotone behaviour i.e. it decrease
from infinity to some minimal radius value $r_0$ at the throat such that $\epsilon(r_0) = r_0=r$, and then again goes to infinity. }
\item{$l(r)$ is an appropriate radial distance which is related to the radial coordinate $r$, given by
\begin{equation}\label{6}
l(r)=\pm \int^{r}_{r_0} \frac{dr}{\sqrt{1-\frac{\epsilon(r)}{r}}},
\end{equation}
implies the condition $1-\frac{\epsilon(r)}{r}\geq0$.}
\item{For spatial geometry, space must be asymptotically flat i.e, $\frac{\epsilon(r)}{r}\rightarrow0$ as $l\rightarrow\pm\infty$.}
\item{In wormhole geometry, at the throat $\epsilon(r_0) = r_0=r$, the flaring out condition must be fulfilled which implies $\frac{\epsilon(r)-r \epsilon'(r)}{\epsilon'(r)}>0$, here $\epsilon'(r)=\frac{d\epsilon}{dr}$. Further, the
condition $\epsilon'(r)<1$ is essential for the wormhole structure.}
\end{itemize}
According to the classical GR, presence of exotic matter in wormhole structure is the main cause of violation of weak energy condition
(WEC) denoted as $T_{\zeta \eta}u^{\zeta} u^{\eta}\geq 0$ for any space-like vector $k^u$ \cite{haw2}. Hochberg and Visser \cite{hoc2,hoc4} provided
the generalization of results with exotic matter already reported by  Morris-Thorne \cite{mor}. It was demonstrated that the wormhole throat does not obey
the NEC. Raychaudhuri equation reads \cite{ray}
\begin{eqnarray}
\frac{d \vartheta}{d \tau} &=& - \frac{\vartheta^2}{3}-R_{\zeta \eta} v^{\zeta} v^{\eta}-\sigma_{\zeta \eta} \sigma^{\zeta \eta}-w_{\zeta \eta} w^{\zeta \eta},\label{aa}\\
\frac{d \vartheta}{d \tau} &=&- \frac{\vartheta^2}{2}-R_{\zeta \eta} u^{\zeta} u^{\eta}-\sigma_{\zeta \eta} \sigma^{\zeta \eta}-w_{\zeta \eta} w^{\zeta \eta}.\\\nonumber
\end{eqnarray}
Here $\vartheta$, $w_{\zeta \eta}$ and $\sigma_{\zeta \eta}$ stands for expansion, rotation and shear of the congurence associated by the vector field $v^{\zeta}$.
 Now as $\sigma^2=\sigma_{\zeta \eta} \sigma^{\zeta \eta}$ and for any hypersurface orthogonal congruences $w_{\zeta \eta}\equiv0$, due to that, the condition
for attractive gravity scale down to $R_{\zeta \eta} v^{\zeta} v^{\eta}\geq0$. After some manipulations GR field equations, one can rewrite the
last expression in terms of energy momentum tensor as $T_{\zeta \eta} v^{\zeta} v^{\eta}\geq0$. The energy conditions named as NEC,
weak energy conditions (WEC)
are described as
 \begin{equation}
NEC:~~~~\rho+ p_{r} \geq 0,~~   \rho +p_{t} \geq 0,~~~~~~~~~~~~~~~~~~~~~~~~~
\end{equation}
\begin{equation}
WEC:~~~\rho \geq0,~~ \rho+ p_{r}\geq 0,~~  \rho +p_{t} \geq 0,~~~~~~~~~~~~~~~~~~~
\end{equation}
respectively. For the formulation of wormhole structure the NEC must be violated due to the exotic matter. Note that the NEC is the weakest condition
as compare to others.

\section{$f(R)$ Gravity}

In this section, we briefly explain the modified $f(R)$ theory of gravity. $f(R)$ gravity starts with the modified form of standard Einstein-Hilbert
action defined as
\begin{equation}\label{1}
\mathcal{S} = \int \sqrt{-g}[\frac{1}{2 \kappa}(f(R))+\mathcal{L}_m] d^4x.
\end{equation}
Here $\kappa= 8 \pi G$, for simplicity we consider $\kappa=1$ for throughout this work. $f(R)$ is a generic algebraic expression of the Ricci scalar $R$ and $\mathcal{L}_m$ the matter of Lagrangian field. One may recover the standard Einstein-Hilbert action by replacing $f(R)$ with $R$.  By varying the above action with respect to the metric $g_{\zeta \eta}$ yields the following field equations:
\begin{equation}\label{2}
F(R)R_{\zeta \eta}-\frac{1}{2}f(R)g_{\zeta \eta}-\nabla_\zeta \nabla_\eta F(R)+g_{\zeta \eta}\square F(R)=8\pi T^{(m)}_{\zeta \eta},
\end{equation}
where $\square =\nabla_\zeta \nabla^\zeta$ with $\nabla_\zeta$ as the covariant derivative. $T^{(m)}_{\zeta \eta}$ is the standard matter energy-momentum
tensor and $F(R)$ is the derivative of $f(R)$ with respect to the Ricci scalar $R$. As we have mentioned above, one may recover the field equation of GR from
Eq.($\ref{2}$) by setting $f(R)=R$. The energy-momentum tensor for anisotropic fluid is defined as :
\begin{equation}\label{3}
{T^{\zeta}_{\eta}}^{(m)}=(\rho + p_t)u^{\zeta} u_{\eta}-p_t g^{\zeta}_{\eta}+(p_r -p_t) v^\zeta v_\eta,
\end{equation}
where $u_{\zeta}$ represent the velocity four vector, with $u^{\zeta} u_{\eta}=-v^\zeta v_\nu=1$. Here the usual energy density is denoted as $\rho$, $p_r$ and $p_t$ is the radial and tangential pressure respectively. We can rewrite the above equation in the standard form of Einstein field equations as
\begin{equation}\label{3}
G_{\zeta \eta}=R_{\zeta \eta}-\frac{1}{2}R g_{\zeta \eta}=T^{(eff)}_{\zeta \eta}=\frac{8 \pi}{F} T^{(m)}_{\zeta \eta}+\frac{1}{F}(\nabla_\zeta \nabla_\eta F(R)-(\square F(R)+ \frac{1}{2} R F(R)-\frac{1}{2} f(R))g_{\zeta \eta}),
\end{equation}
Now using Eqn. ($\ref{3}$) along with Eqn. ($\ref{4}$), one can find the following field equations as \cite{lob}

\begin{eqnarray}
\rho &=& \frac{F \epsilon'}{r^2},\label{5d}\\
p_r &=& \frac{-F \epsilon}{r^3} + \frac{F'(r \epsilon'-\epsilon)}{2 r^2} - F''(1- \frac{\epsilon}{r}),\label{5e}\\
p_t &=& \frac{-F'}{r} (1- \frac{\epsilon}{r}) + \frac{F}{2 r^3} (\epsilon- r \epsilon').\label{5f}
\end{eqnarray}

Bronnikov et al. \cite{bro} discussed the stability and ghost-free scalar-tensor
phantom for static traversable wormhole solutions. Further, Bronnikov et al. \cite{bro} demonstrated the wormhole existence in the context of scalar-tensor and $f(R)$ gravity. It was concluded that non-existence theorem could be violated when $f(\phi)$ or equivalently $\frac{df}{dR}=F(R)$ is negative. This also implies that for $f(\Phi),~ F(R)~>0$, there does not exist a static wormhole that satisfy the NEC.

Due to the problematic nature of exotic matter, it is suggested to minimize the presence of exotic matter to make it appropriate for travelling \cite{5har1, 5li, lob}. In classical GR the wormhole geometry is supported by the presence of exotic matter that violates the NEC and WEC. Whereas, in modified theories the scenario may be different. Specifically in $f(R)$ gravity, it is possible to find the wormhole solutions that obey the energy conditions due to its higher order curvature terms \cite{lob}. One can observe from Eqs. ($\ref{5d}$) and ($\ref{5e}$)

\begin{eqnarray}
\rho + p_r \mid_{r=r_0} >0~~~ if~~~  F(r \epsilon'-\epsilon)\mid_{r=r_0} +  F' (r \epsilon'-\epsilon) \mid_{r=r_0} >0.
\end{eqnarray}

In wormhole geometry, at the throat $\epsilon(r_0) = r_0=r$, the flaring out condition $\frac{\epsilon(r)-r \epsilon'(r)}{\epsilon'(r)}>0$ is a fundamental property. We can also take its negative as $r \epsilon'(r)-\epsilon(r)<0$ and choose the parameters that formulate the violation of non-existence theorem $F=\frac{df}{dR} <0$ at the throat. By imposing these conditions on $\rho+ p_r >0$, one can find the key point that if $F' <0$ at $r=r_0$, we get a wormhole solution that obeys the NEC at the throat in the context of logarithmic corrections. It is clear from Eqn. ($\ref{5d}$) that the particular solution in the context of $f(R)$ gravity that violates the non-existence theorem can also obey WEC if $\epsilon'<0$.

\subsection{Logarithmic-Corrected $R^2$ Gravity}

It is possible to develop fruitful models of inflation with the help of eminent Starobinsky model \cite{star}. In present paper, we are using $R^2$
Starobinsky inflation model, which may clarify both the early-time and the late-time acceleration cycles . In constant-roll inflation, use of
logarithmic corrections helps to obtain observational indices which are compatible with the latest Planck data. Particularly, the constant-roll condition
enlarges the space of parameters, and it helps to make it feasibly compatible with the observational constraint. Our main focus is
to check the energy conditions test and violation of non-existence theorem for static spherically wormholes by using this model defined by Elizalde et al. \cite{eliz}
\begin{equation}\label{7}
f(R)=R+ \alpha R^2+ \beta R^2 Log(\beta R),
\end{equation}
where $\alpha$ and $\beta$ are constant free parameters with some appropriate dimensions. It has been demonstrated that this model may provide compatibility with the latest observational data \cite{eliz}.
In particular, due to fact that logarithmic corrections are induced by one-loop effects in quantum gravity, there is still much interest to inflationary phenomenology in the context of logarithmic corrected modified
gravity \cite{104,105}. Moreover, models may become ghost-free and the cosmological solutions may correspond inflation when the model parameters $\alpha$ and $\beta$ are both positive. The cosmological solutions in the context of $f(R)$ theories of gravity may justify the existence of  accelerating
universe when $\beta<0$  \cite{noj1}. However, with negative $\beta$ one has to restrict with negative curvature scenarios to obtain a real logarithmic value.
It has been shown that some wormholes can be described by negative curvature and such wormholes have
throat sections in the form of tori and are traversable and stable in the cosmological context \cite{106}.

As mentioned in key properties that for the existence of traversable wormhole, the red-shift parameter
$\psi$ must be finite everywhere. Moreover, the conditions $\psi=d\rightarrow0$ as $r\rightarrow\infty$ suggested that $\psi$ should be asymptotically flat as
well. To preserve these conditions we fix the red-shift parameter as a constant function i.e, $\psi=d$. Now, after substituting Eqs.($\ref{6}$) and ($\ref{7}$)
along with assumption $\psi=d$ in Eqs. ($\ref{5d}$)-($\ref{5f}$), we get the following set of equations
\begin{eqnarray}
&& \rho = \frac{\epsilon'}{r^4} \bigg[r^2 +4(\alpha+ \beta^2 + \beta Log[\frac{2 \beta \epsilon'}{r^2}]) \epsilon'  \bigg],\label{a}\\
&& p_r =\frac{1}{r^5 \epsilon'} \bigg[-\epsilon~\epsilon' (r^2 +4(\alpha+ \beta^2 + \beta Log[\frac{2 \beta \epsilon'}{r^2}]) \epsilon')+ 2(\alpha+ \beta+ \beta^2 + \beta Log[\frac{2 \beta \epsilon'}{r^2}]) \epsilon' (\epsilon- r \epsilon') (2 \epsilon'- r \epsilon'')-4 (r - \epsilon)
\nonumber\\
&&  \bigg(2 (3 \alpha +  \beta (5+ 3 \beta)+ 3 \beta Log[\frac{2 \beta \epsilon'}{r^2}]) \epsilon'^2 + r^2 \beta~ \epsilon''^2 + r~ \epsilon' (-4(\alpha + \beta (2+ \beta)+ \beta Log [\frac{2 \beta \epsilon'}{r^2}]) \epsilon'') + r (\alpha + \beta + \beta^2 +
\nonumber\\
&& \beta Log [\frac{2 \beta \epsilon'}{r^2}]~ \epsilon''' ) \bigg)        \bigg],\label{b}\\
&& p_t = \frac{1}{2 r^5} \bigg[(\epsilon- r \epsilon') (r^2 + 4 (\alpha + \beta^2 + \beta Log [\frac{2 \beta \epsilon'}{r^2}])\epsilon' ) -8(\alpha + \beta +\beta^2 + \beta Log [\frac{2 \beta \epsilon'}{r^2}] )
 (r - \epsilon) (-2 \epsilon' + r \epsilon'' )  \bigg] .\label{c}
\end{eqnarray}

In the following work, we are interested for solutions where $\frac{df}{dR} <0$, to get the wormhole geometry that violates the non-existence theorem. In next section, we have shown the region plots for $\frac{df}{dR} <0$, according to the shape functions and discussed NEC and WEC.

\section{Numerical Solutions}
In this section, our main focus is to discuss the evolution of static wormholes by using specific choices of shape functions for anisotropic fluid.
Further, we discuss the numerical solutions of shape function for perfect fluid and barotropic fluid. It is worthy to mention here that we analyze $\rho$ ($g/cm^3$), $p_r$ and $p_t$ ($dyne/cm^2$)  units with radial coordinate $r$ in kilometers [Km].

\subsection{Anisotropic Fluid}

\begin{figure}\center
\begin{tabular}{cccc}
\epsfig{file=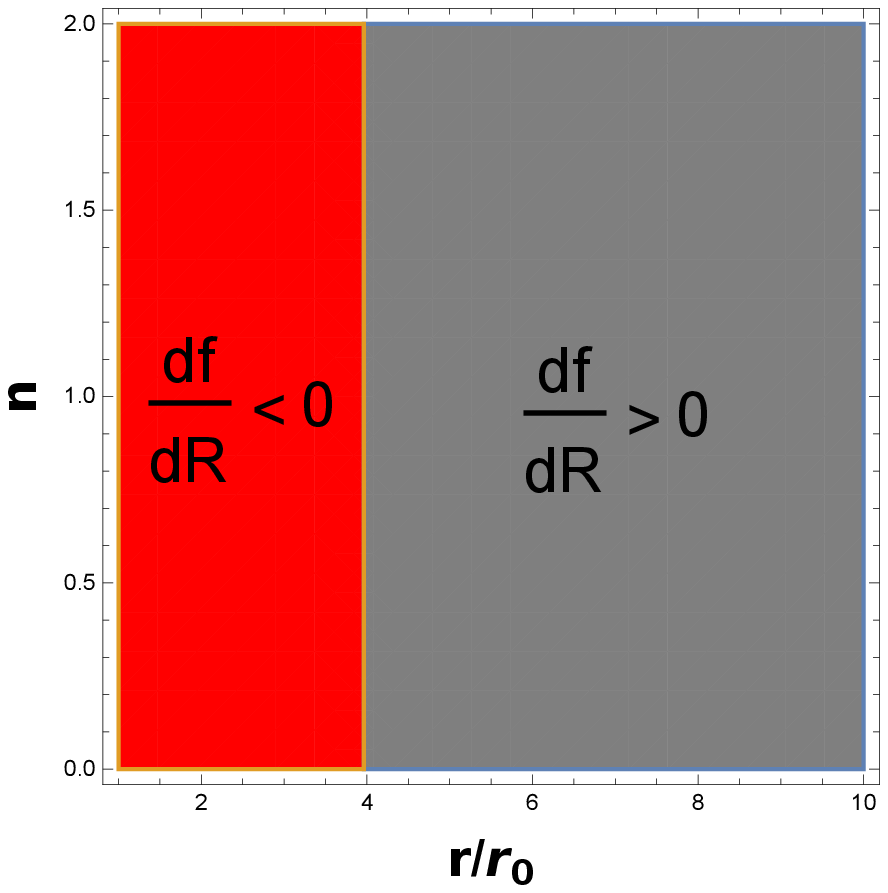,width=0.27\linewidth} &
\epsfig{file=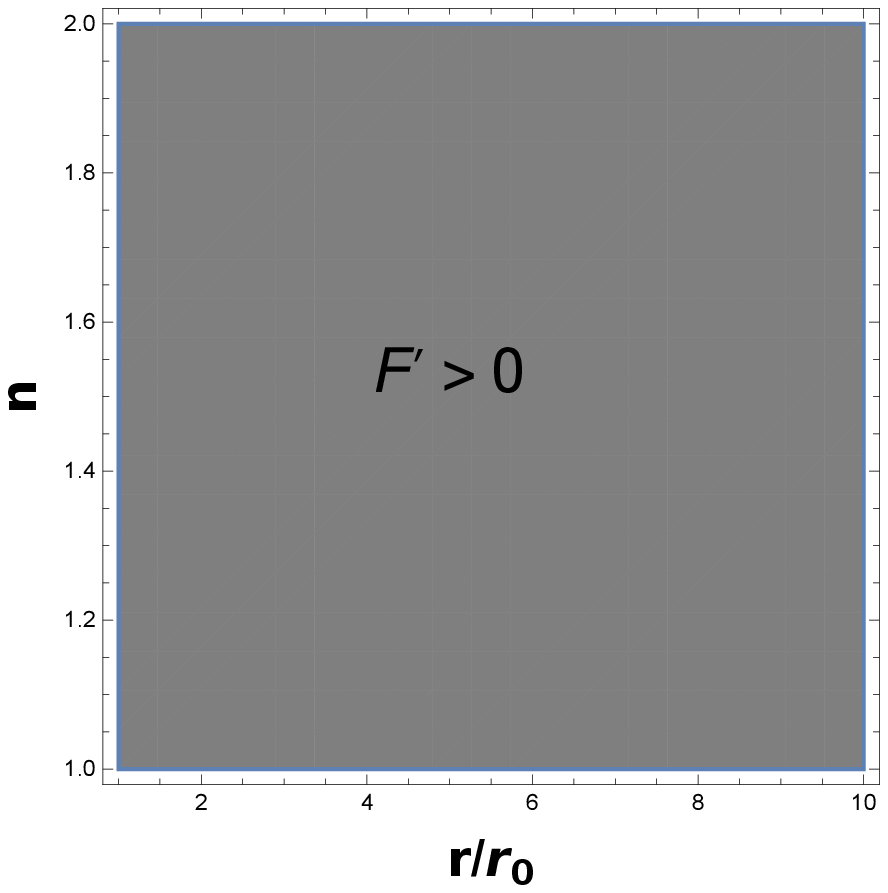,width=0.27\linewidth} \\
\end{tabular}
\caption{Red Region shows that $F=\frac{df}{dR}<0$ (left side) and gray region shows $F'>0$ (right side), for shape function $\epsilon(r)=(r_0)^{n+1} r^{-n}$ with $r_0=2$ Km, $\alpha=20,~~ \beta =-0.1$ and $0 \leq n\leq 2$. So, we get the violation of nonexistence theorem at the throat and $F'>0$ provides the wormhole geometry which is filled with exotic matter at the throat.}\center
\label{Fig:51}
\end{figure}

\begin{figure}\center
\begin{tabular}{cccc}
\epsfig{file=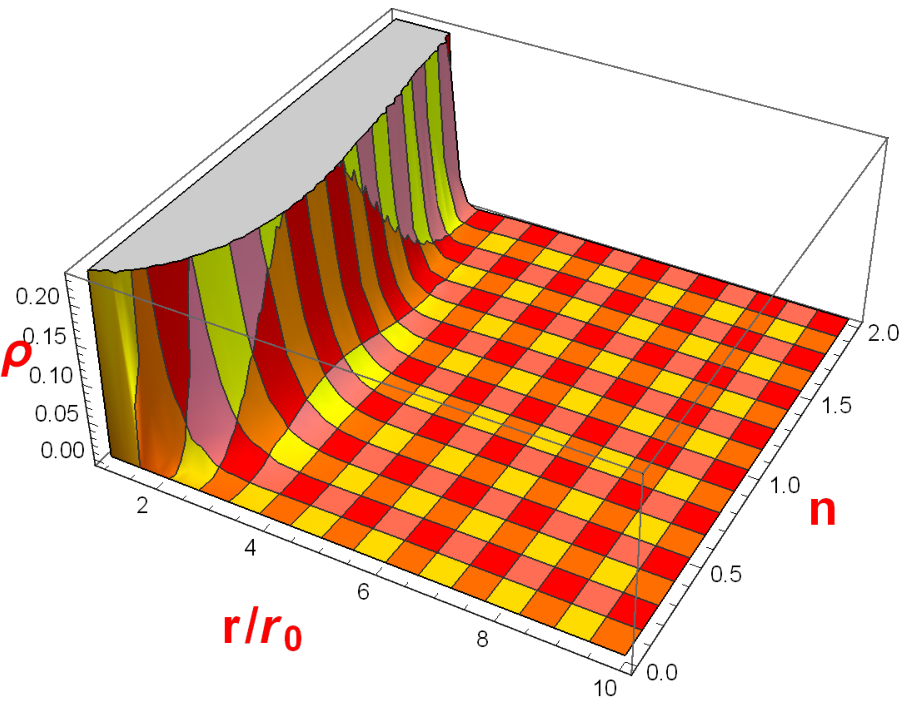,width=0.27\linewidth} &
\epsfig{file=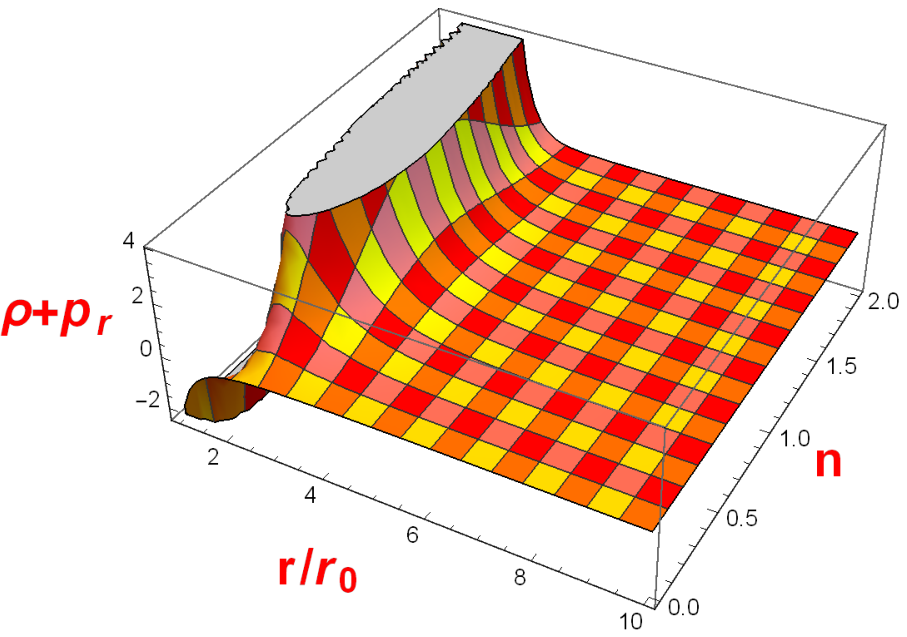,width=0.29\linewidth} &
\epsfig{file=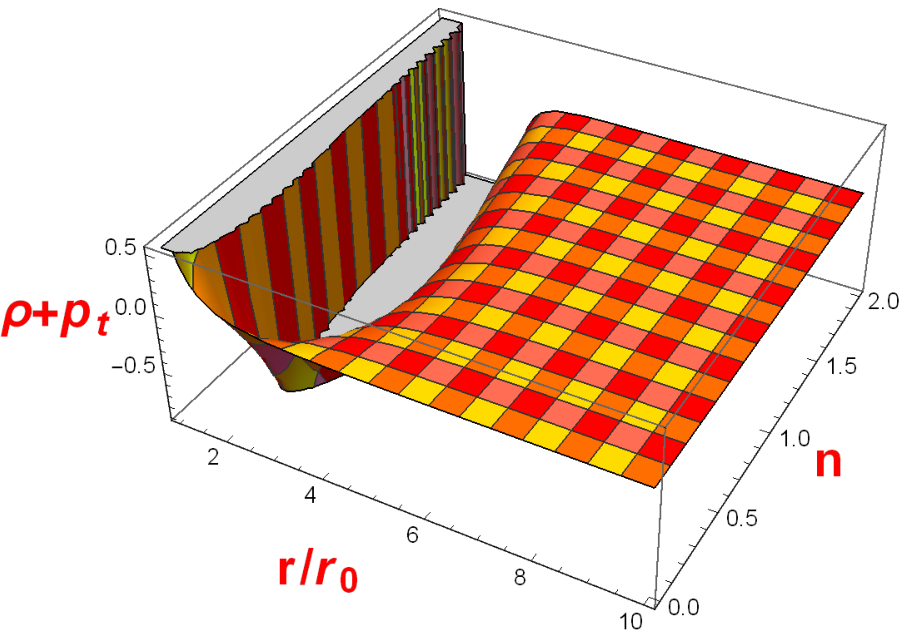,width=0.29\linewidth} \\
\epsfig{file=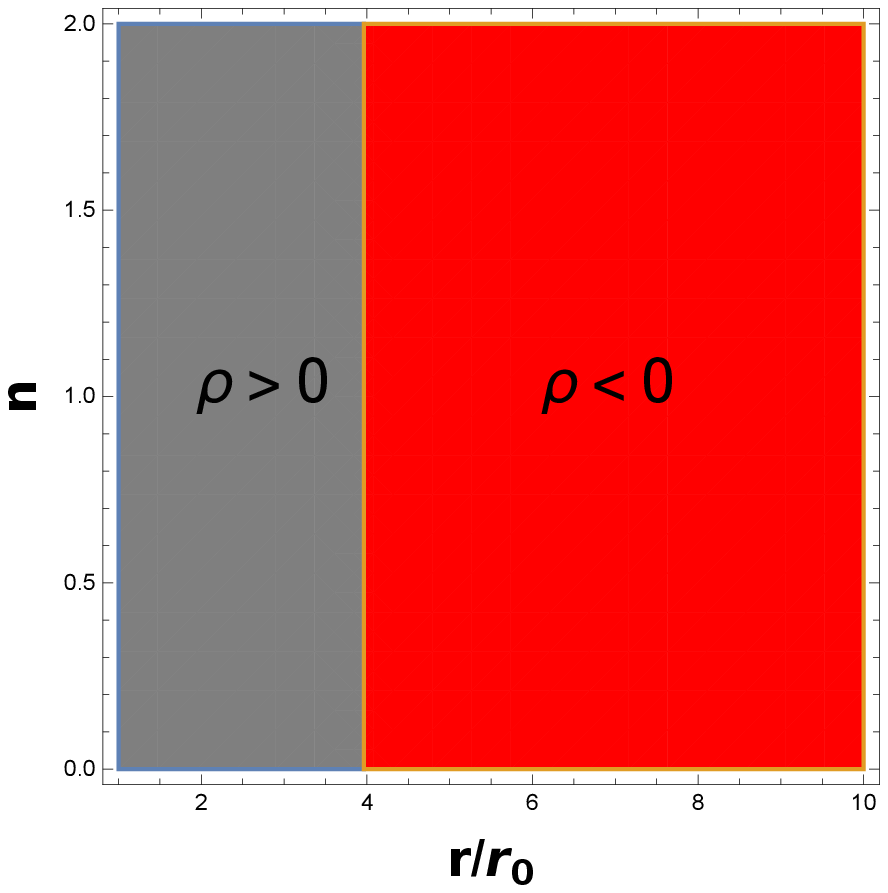,width=0.29\linewidth, height=1.5in} &
\epsfig{file=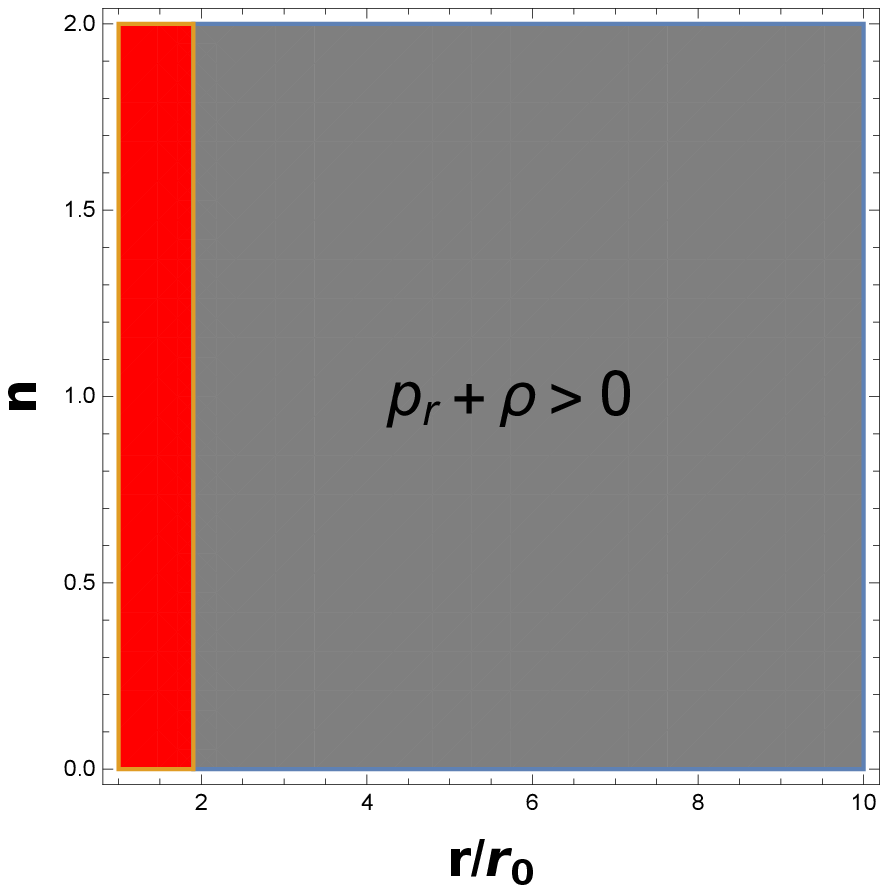,width=0.29\linewidth, height=1.5in} &
\epsfig{file=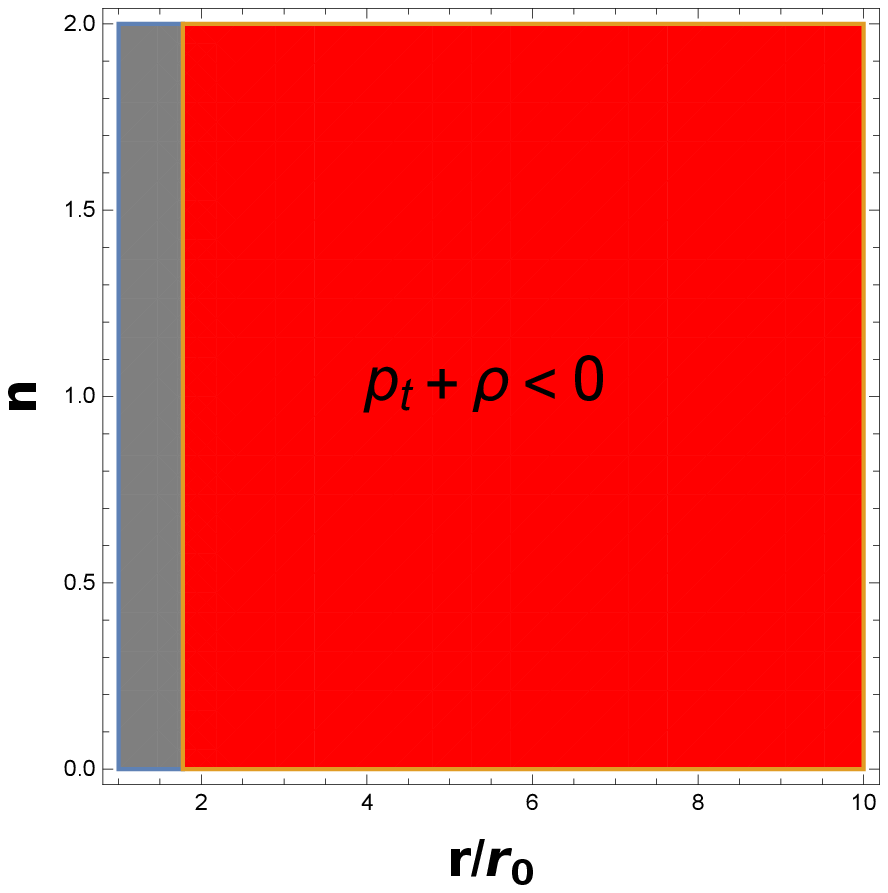,width=0.29\linewidth, height=1.5in}\\
\end{tabular}
\caption{For shape function $\epsilon(r)=(r_0)^{n+1} r^{-n}$ with $r_0=2$ Km, $\alpha=20,~~ \beta =-0.1$ and $0 \leq n\leq 2$. Above three plots shows the behaviour of $\rho$, $\rho + p_r$ and $\rho+p_t$, and below three plots are their corresponding regions. It is clear that at the throat $\rho <0$, $\rho + p_r <0$ and $\rho + p_t >0$, but outside the throat $\rho >0$, $\rho + p_r >0$ and $\rho + p_t <0$. It shows the presence of exotic matter in wormhole geometry.}\center
\label{Fig:52}
\end{figure}
The investigation of spherically symmetric traversable wormhole
spacetimes has been mostly focused in matter sources
with anisotropic pressures. For anisotropic case, different researchers use various types of shape functions to discuss static wormholes.
Firstly, we use the following shape function \cite{jam,shar2,sham3}
\begin{equation}\label{8}
\epsilon(r)=(r_0)^{n+1} r^{-n},
\end{equation}
where $n$ and  $r_0$ are arbitrary constants. One can observe that for different choices of $n$, we can get different shape parameters. So many authors use
different values for $r$ as per requirement. For example Zubair et al. \cite{zub2} investigate the static
spherically symmetric wormhole evolution in detail for $n=1/2$ in $f(R,T)$ gravity while Pavlovic and Sossich \cite{pav} use same value to discuss wormholes in $f(R)$ theory of gravity. On the other hand, Lobo and Oliveira \cite{lob} discussed the wormhole structure in $f(R)$ gravity for $n=1$ and $-1/2$. As mentioned above that for spatial geometry, space must be asymptotically flat i.e, $\frac{\epsilon(r)}{r}\rightarrow0$ as $l\rightarrow\infty$. The shape function Eqn ($\ref{8}$) fulfills this condition.

The involvement of free parameters in $f(R)$ models has a momentous role in the question of existence of wormhole.
The wormhole solution corresponding to the parameters $\alpha=20,~~ \beta =-0.2$ and $0 \leq n\leq 2$ is given in Figs. $\ref{Fig:51}$ and $\ref{Fig:52}$.
For the violation of non-existence theorem, $\frac{df}{dR}$ must be negative. In Fig. $\ref{Fig:51}$, we have plotted the region for $\frac{df}{dR} <0$ corresponding to the shape function Eq. ($\ref{8}$) with $r_0=2$ Km. It is clear from Fig. $\ref{Fig:51}$, $\frac{df}{dR} <0$ at the throat and shows the existence of static spherically symmetric wormhole geometry. In Fig. $\ref{Fig:52}$, we evaluate the corresponding behaviours of $\rho,~~ \rho + p_r$ and $ \rho+ p_t$ along with their region plots. It can be seen that the condition $\rho + p_r$ is violated at the throat. We also observe that any combination of free parameters that violates the non-existence theorem, violates the WEC. Thus, we can say that one cannot avoid the presence of exotic matter for static spherically symmetric wormhole for shape function Eqn ($\ref{8}$) in the context of logarithmic corrected $f(R)$ model.

Similarly, Samanta et al. \cite{sam} introduced exponential shape function as
\begin{equation}\label{9}
\epsilon(r)= \frac{r}{e^{(r-r_0)}}.
\end{equation}
\begin{figure}\center
\begin{tabular}{cccc}
\epsfig{file=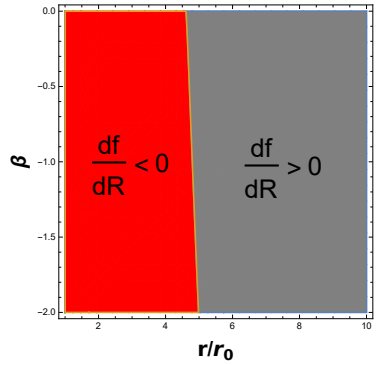,width=0.27\linewidth} \\
\end{tabular}
\caption{Region plot $F=\frac{df}{dR}<0$, for shape function $\epsilon(r)=\frac{r}{e^{(r-r_0)}}$ with $r_0=2$ Km, $\alpha=20$, and $-2 \leq \beta \leq 0$. For this particular shape function we get $F'=0$ for $\beta <0$. So, the positivity of NEC totally depends on the $F=\frac{df}{dR}>0$. }\center
\label{Fig:53}
\end{figure}
\begin{figure}\center
\begin{tabular}{cccc}
\epsfig{file=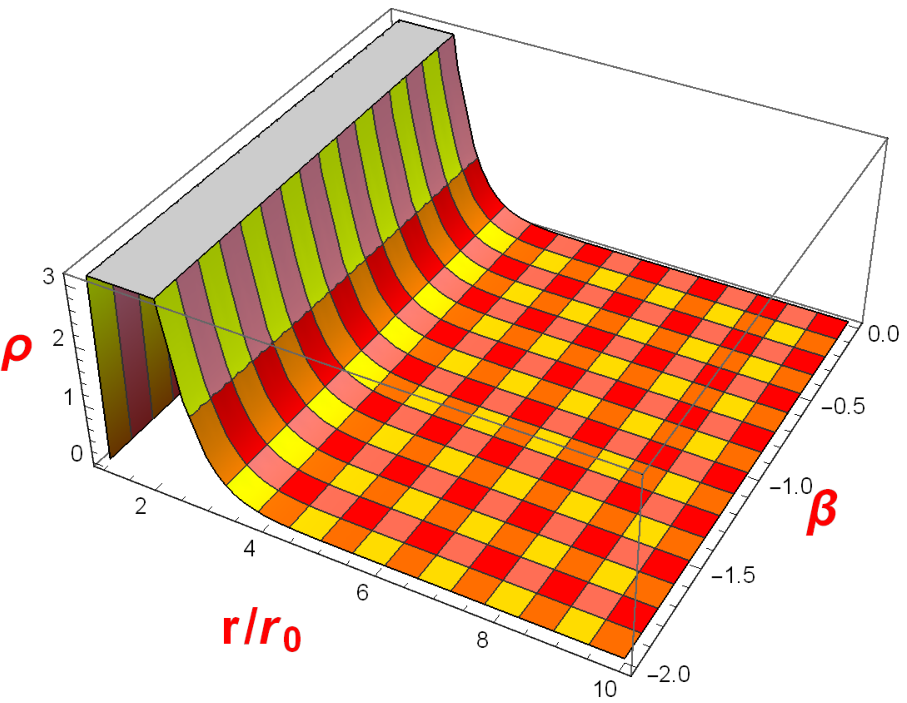,width=0.27\linewidth} &
\epsfig{file=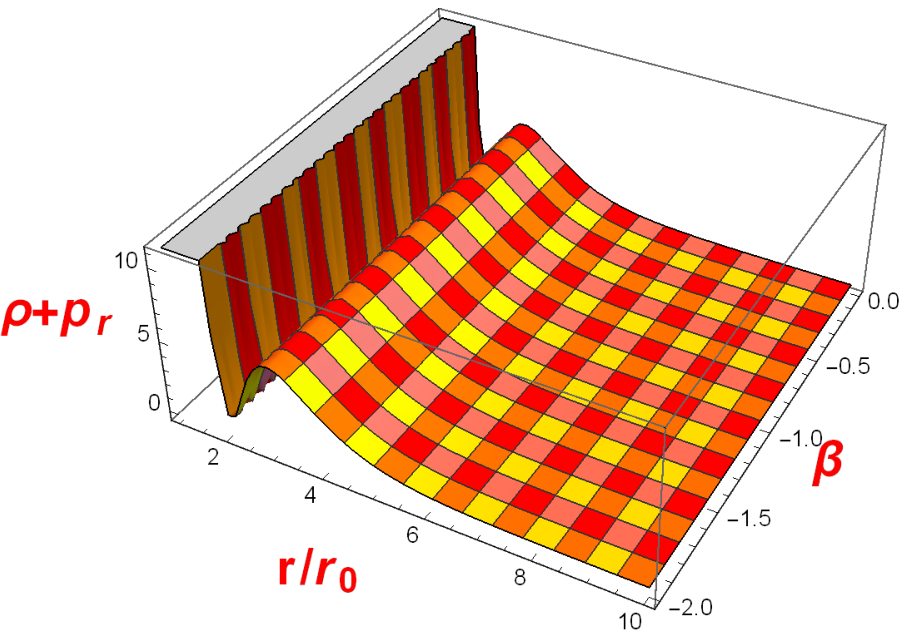,width=0.29\linewidth} &
\epsfig{file=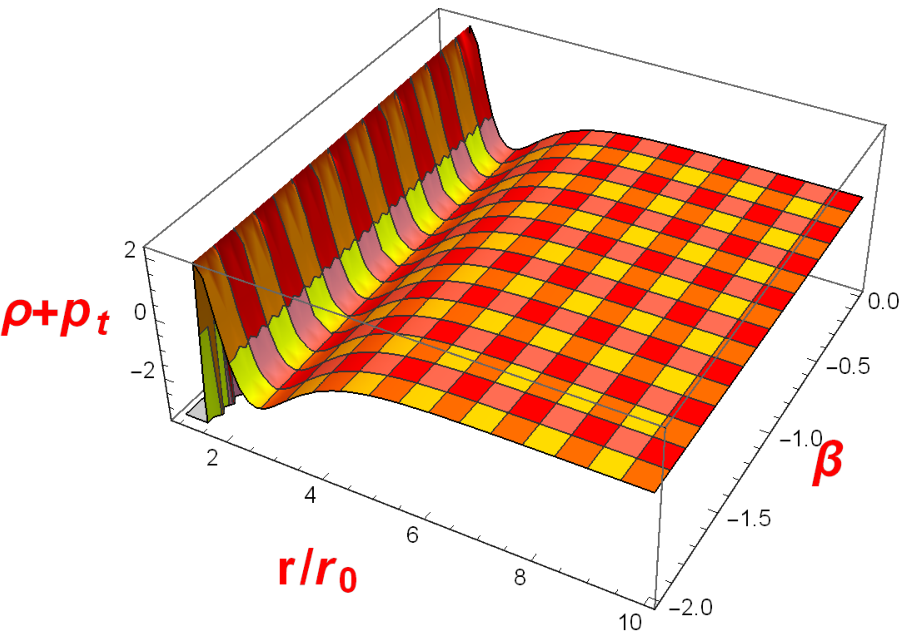,width=0.29\linewidth} \\
\epsfig{file=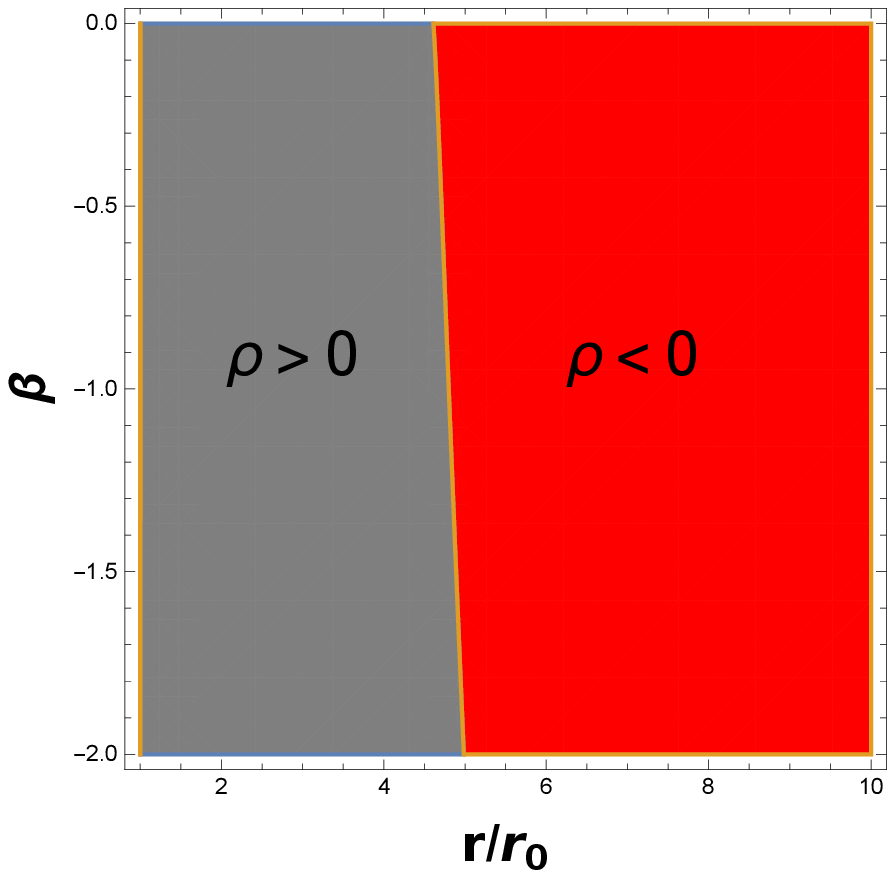,width=0.29\linewidth, height=1.5in} &
\epsfig{file=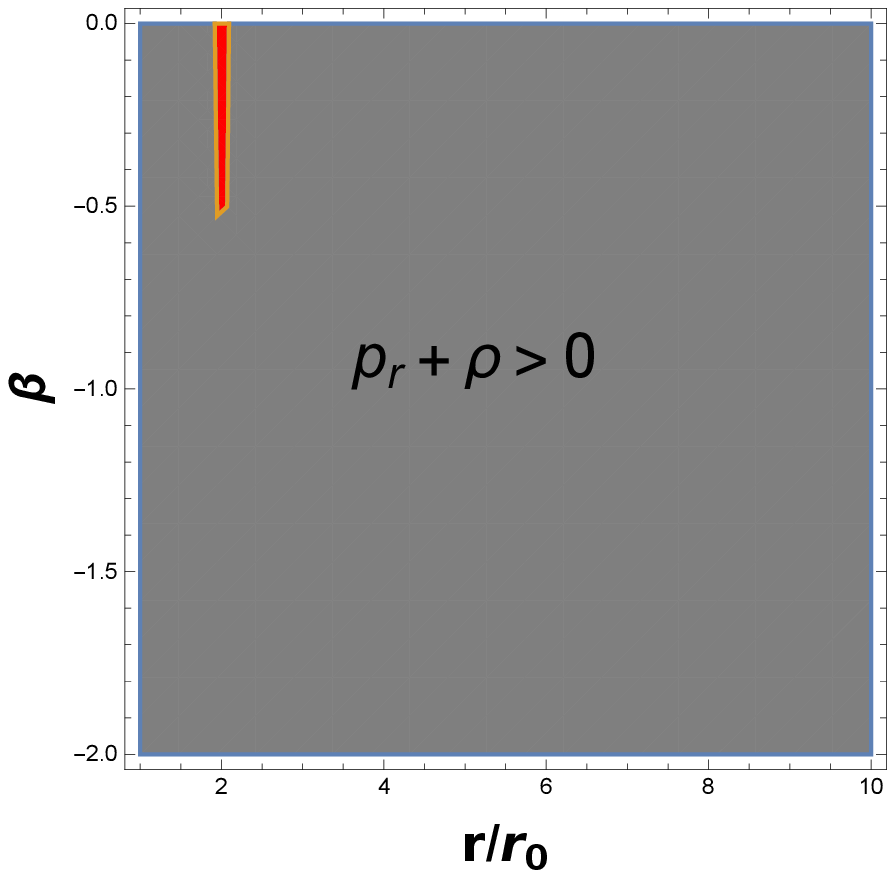,width=0.29\linewidth, height=1.5in} &
\epsfig{file=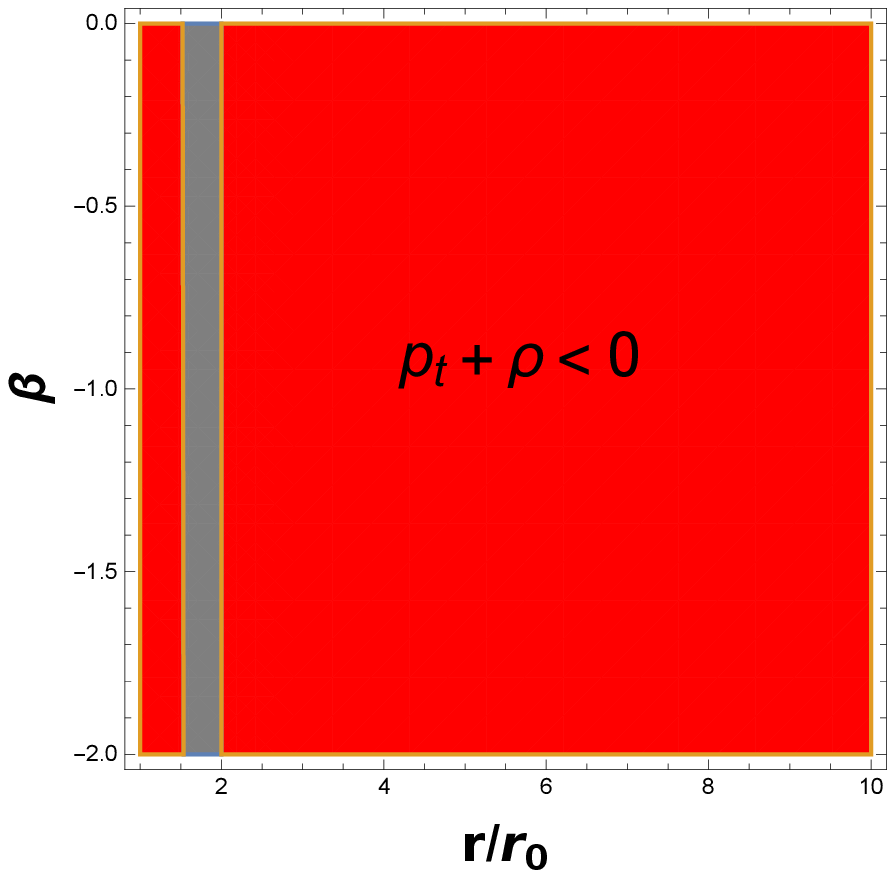,width=0.29\linewidth, height=1.5in}\\
\end{tabular}
\caption{Exponential shape function $\frac{r}{e^{(r-r_0)}}$ with $r_0 =2$ Km, $\alpha=20$, and $-2 \leq \beta \leq 0$. First row shows the evaluation of $\rho$, $\rho + p_r$
and $\rho+p_t$. 2nd row shows their corresponding regions, where $\rho > 0$  $\rho+p_r >0$ and  $\rho+p_t<0$ at the throat.}\center
\label{Fig:54}
\end{figure}
They discussed its feasibility in GR and modified $f(R)$ and $f(R,T)$ theories as well. Firstly, by using this shape function, we found the feasible region where $\frac{df}{dR} <0$ for the parameters $\alpha=20$ and $-2 \leq \beta \leq 0$. Here we set $r_0=2$ Km and it can be seen from Figs $\ref{Fig:53}$ and $\ref{Fig:54}$ that $\frac{df}{dR} <0$, $\rho >0$, $\rho + p_r >0$ at the throat and $\rho + p_t >0$ near the throat. It is also observed that as we increase the value of parameter $\alpha$, our wormhole solution avoids exotic matter (or avoid violation of WEC and NEC) and provides more appropriate solution of wormhole geometry. Due to the problematic nature of exotic matter, it is suggested to minimize the presence of exotic matter to make it appropriate for travelling \cite{5har1, 5li, lob}. We observe that exponential function violates the non-existence theorem and supports the wormhole formulation with less amount of exotic matter with the higher values of parameter $\alpha$.

\begin{figure}\center
\begin{tabular}{cccc}
\epsfig{file=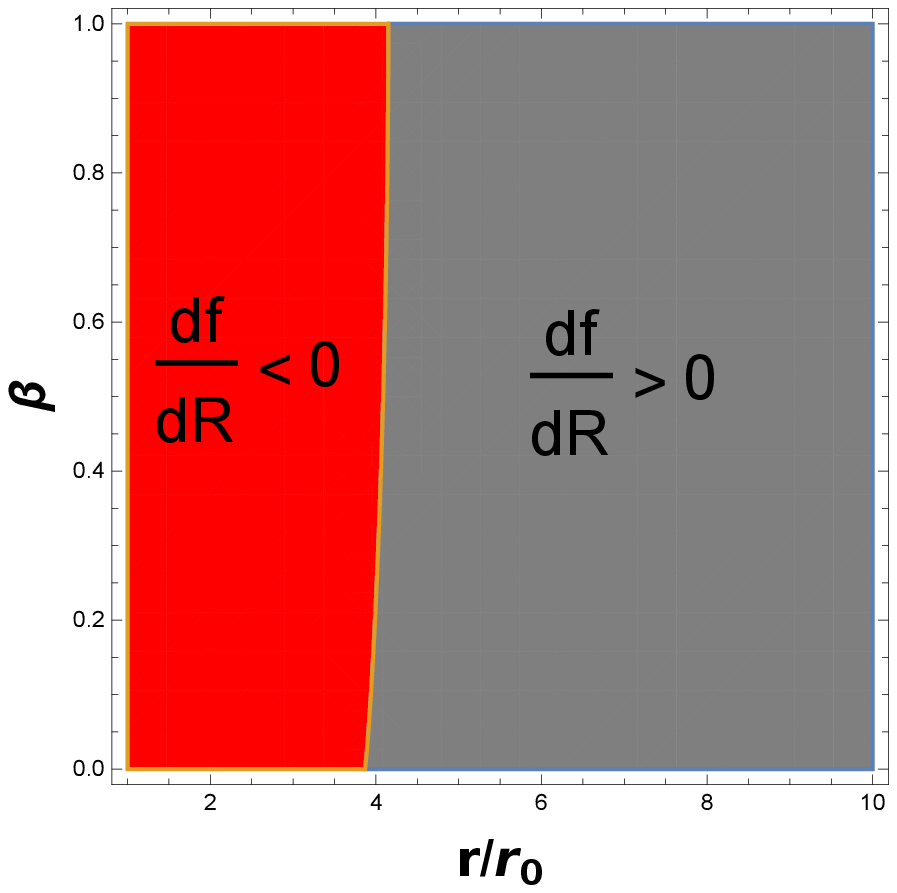,width=0.27\linewidth} &
\epsfig{file=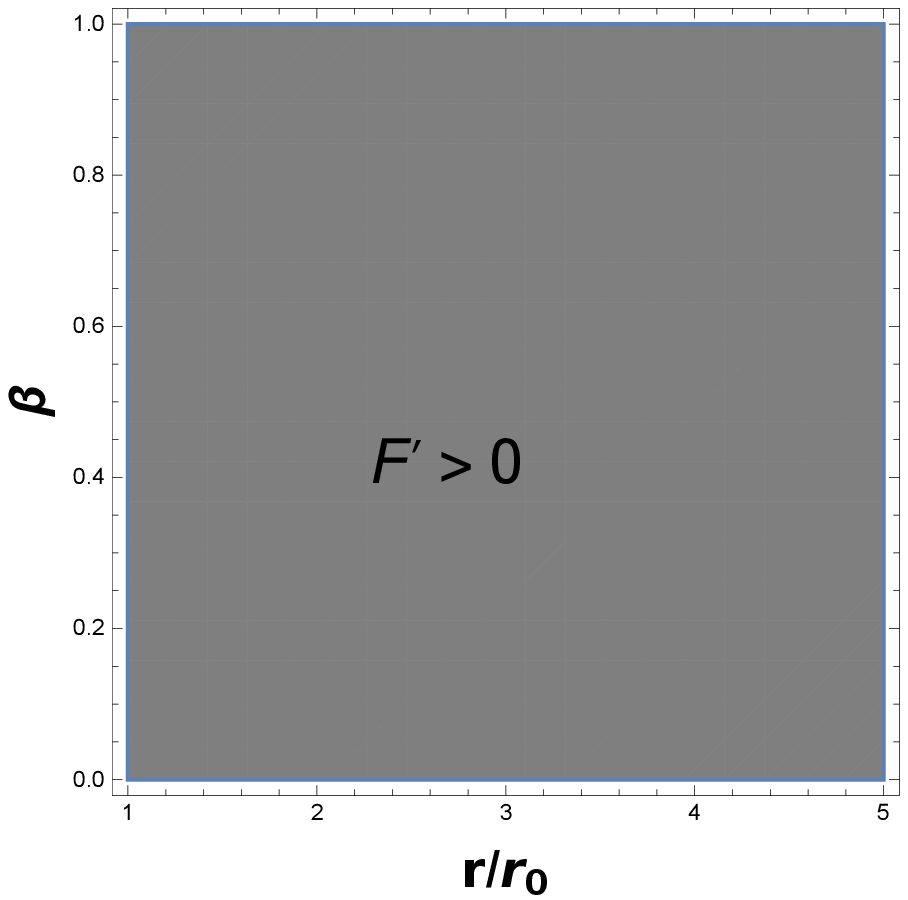,width=0.27\linewidth} \\
\end{tabular}
\caption{Red Region shows that $F=\frac{df}{dR}<0$ (left side) and gray region shows $F'>0$, for shape function $\epsilon(r)= \frac{r_0 Log[1+r]}{Log[1+r_0]}$ at the throat $r_0=2$ Km with $\alpha=-10$ and $0 \leq \beta \leq 1$.}\center
\label{Fig:56}
\end{figure}

\begin{figure}\center
\begin{tabular}{cccc}
\epsfig{file=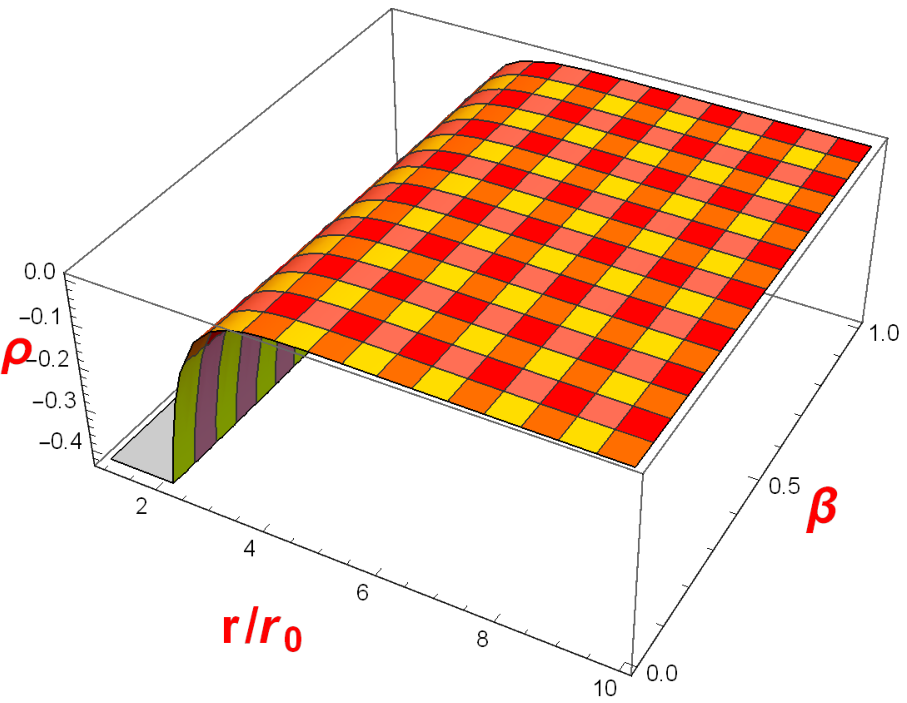,width=0.27\linewidth} &
\epsfig{file=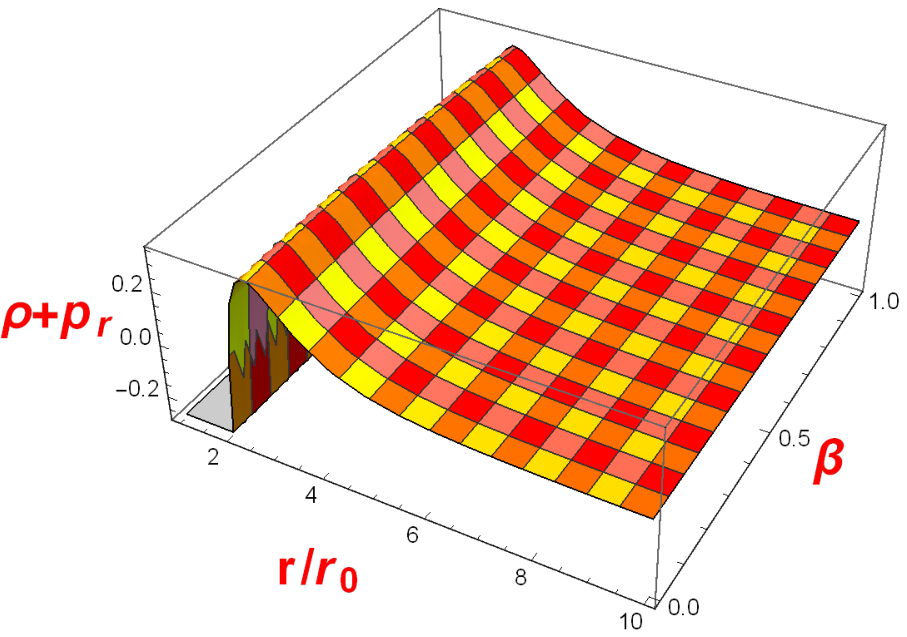,width=0.29\linewidth} &
\epsfig{file=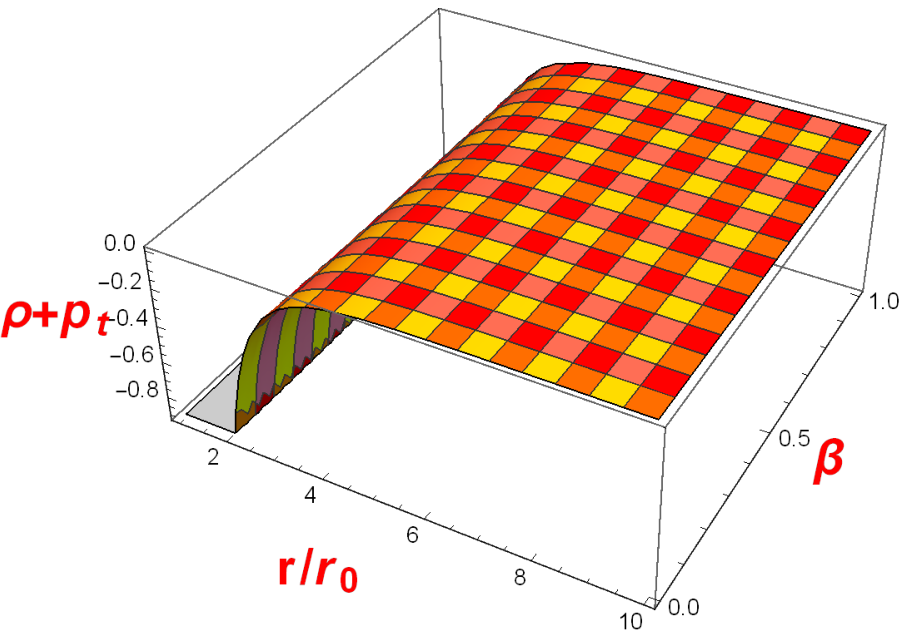,width=0.29\linewidth} \\
\epsfig{file=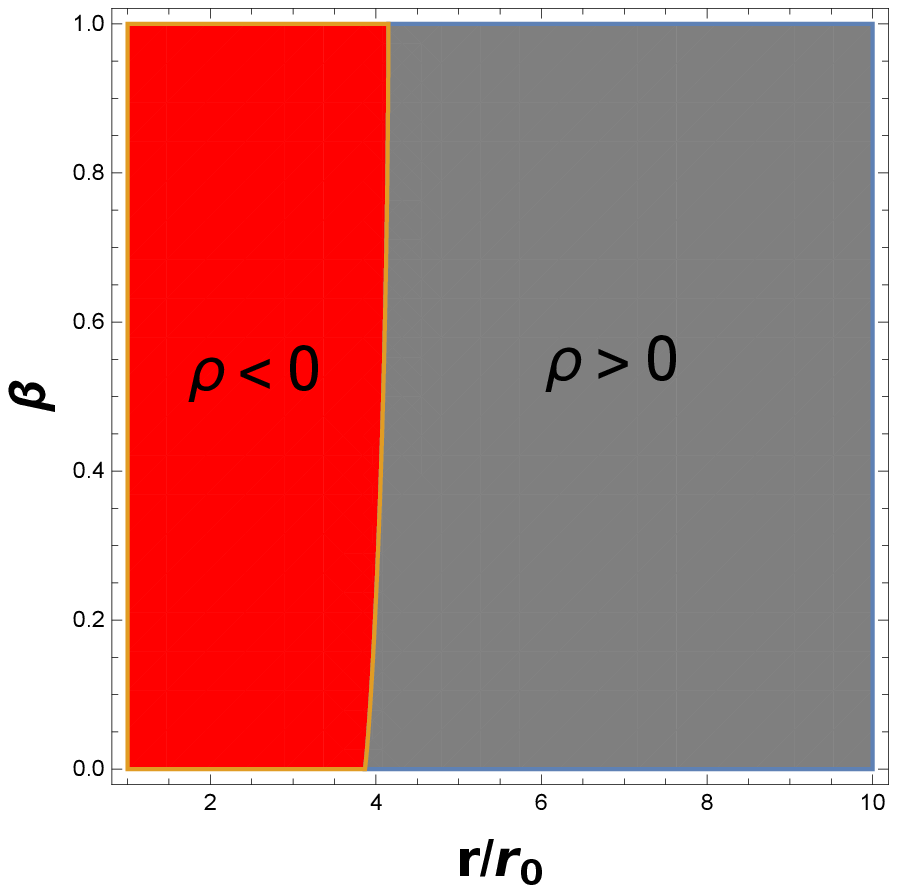,width=0.29\linewidth, height=1.5in} &
\epsfig{file=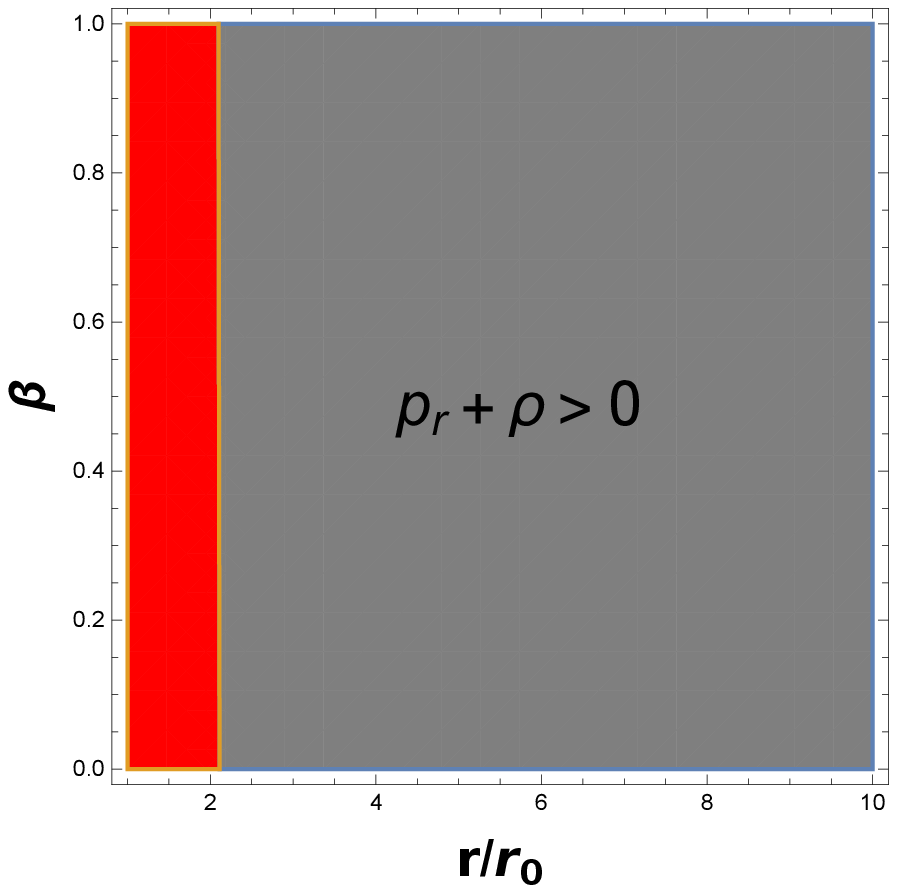,width=0.29\linewidth, height=1.5in} &
\epsfig{file=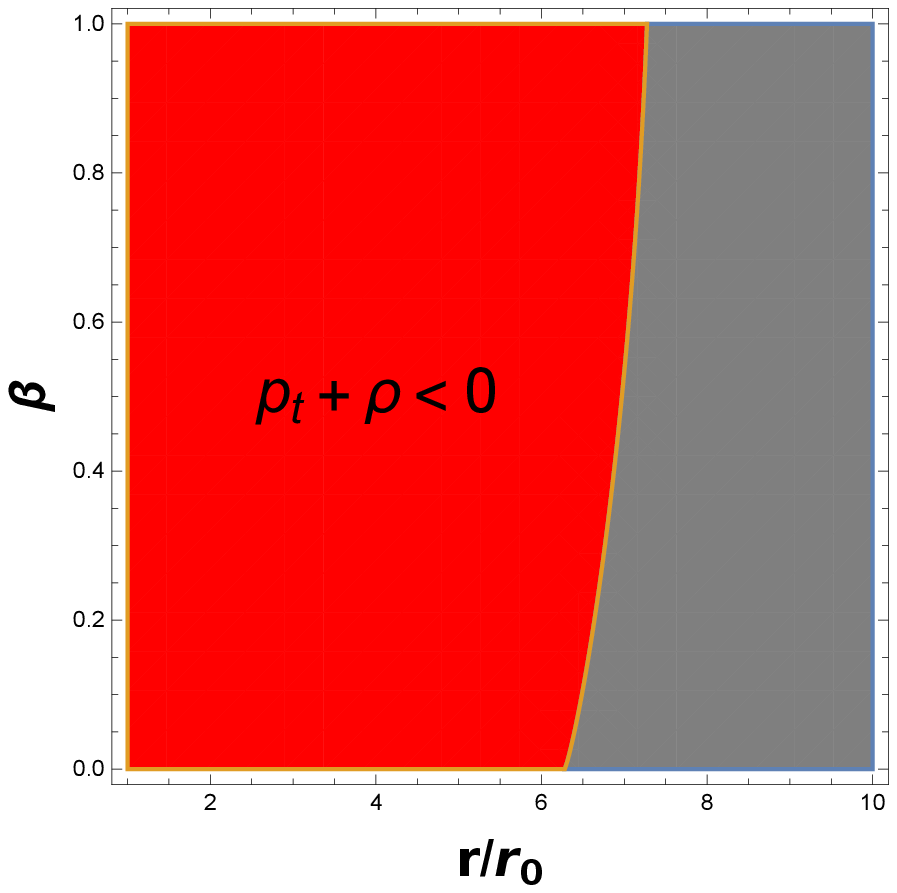,width=0.29\linewidth, height=1.5in}\\
\end{tabular}
\caption{Exponential shape function $\epsilon(r)= \frac{r_0 Log[1+r]}{Log[1+r_0]}$ with $r_0 =2$ Km, $\alpha=-10$, and $0 \leq \beta \leq 1$. First row shows the evaluation of $\rho$, $\rho + p_r$
and $\rho+p_t$. 2nd row shows the feasible regions, where $\rho > 0$  $\rho+p_r >0$ and  $\rho+p_t<0$ at the throat.}\center
\label{Fig:566}
\end{figure}

Recently, Godani and Samanta \cite{5goda} introduced new shape function as

\begin{equation}\label{59}
\epsilon(r)= \frac{r_0 Log[1+r]}{Log[1+r_0]},
\end{equation}
where $r_0$ is a throat radius. The red region in Fig. $\ref{Fig:56}$ shows that $\frac{df}{dR} <0$ at the throat $r_0=2$ Km. Fig. $\ref{Fig:566}$ have shown the violation of WEC and NEC at the throat which indicate that geometry of wormhole fills with exotic matter at the throat. For this analysis we set $\alpha=-10$ and $0 \leq \beta \leq 1$. Further, we found that as we decrease the value of $\alpha$, wormhole solution has shown violation of WEC and NEC in more of the region of wormhole space and increase in $\alpha$ provides $df/dR >0$ i.e. there is no realistic wormhole.
So, we conclude that by using $\epsilon(r)= \frac{r_0 Log[1+r]}{Log[1+r_0]}$ with the logarithmic corrections, one can not found the static spherically symmetric wormhole solutions without exotic matter which satisfies the non-existence theorem for any combination of free parameters.

\subsection{Isotropic Fluid}

\begin{figure}\center
\begin{tabular}{cccc}
\epsfig{file=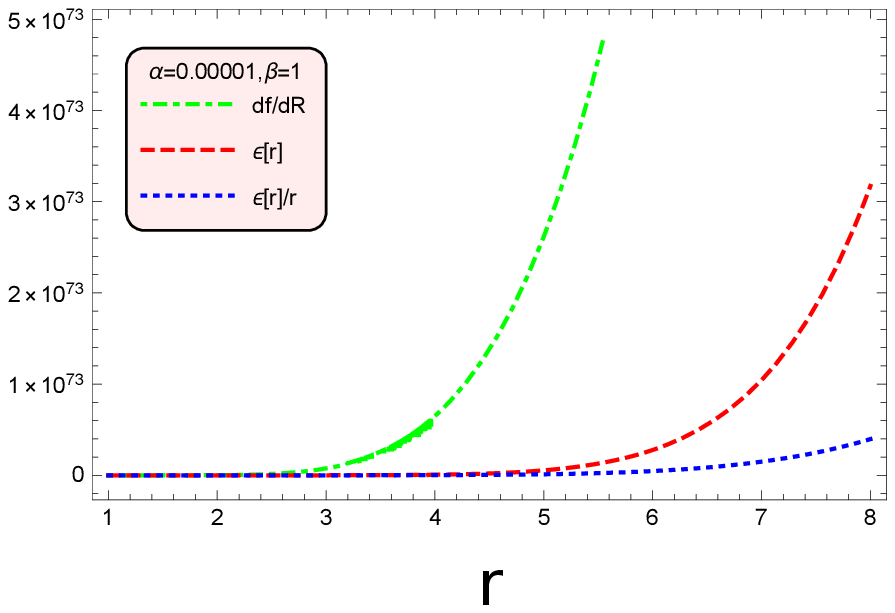,width=0.34\linewidth} &
\epsfig{file=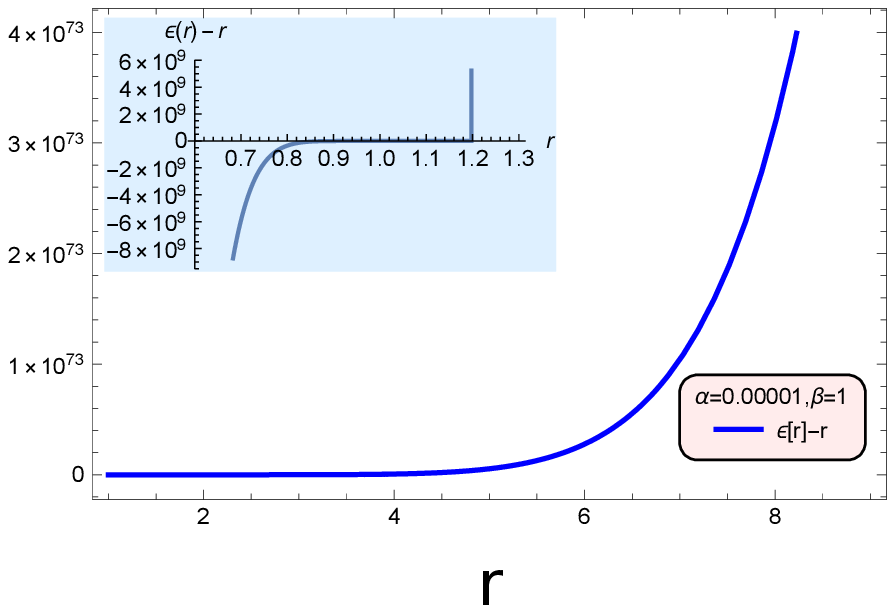,width=0.34\linewidth} &
\epsfig{file=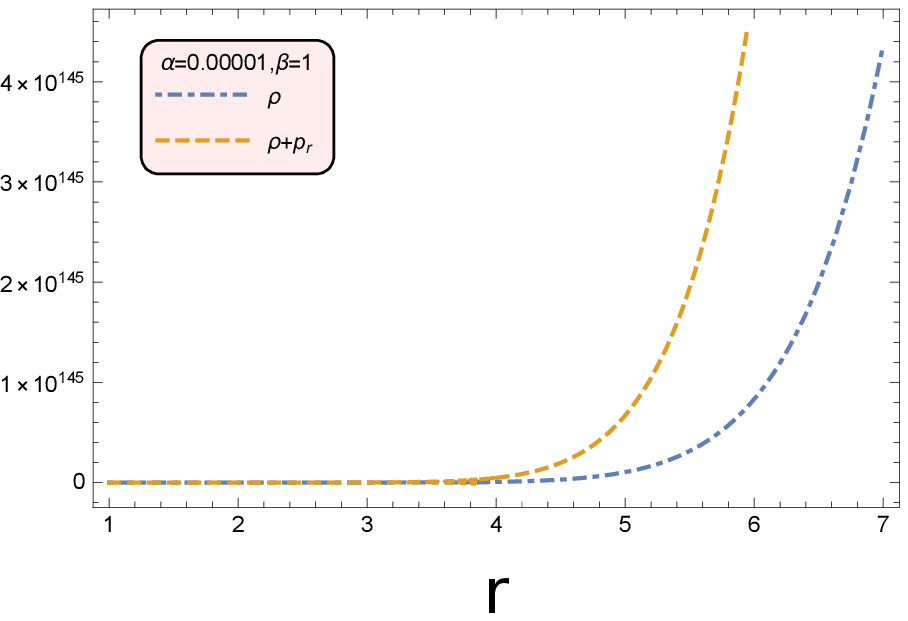,width=0.34\linewidth} \\
\end{tabular}
\caption{First plot shows the behaviour of $df/dR$, $\epsilon$ and $\frac{\epsilon}{r}$ by using $\alpha=0.00001~~ \beta=1$ for isotropic pressure. Second plot verify the position of throat at $r_0 = 1.2$ Km and third plot shows that NEC is respected for wormhole geometry. Thus, we get wormhole solution without exotic matter for isotropic pressure.}\center
\label{Fig:55}
\end{figure}

It is clear that for isotropic fluid $p_r = p_t$. By imposing this condition on Eqs. ($\ref{b}$) and ($\ref{c}$), we get
\begin{eqnarray}
&& 0 =\frac{1}{2 r^5} \bigg[ \frac{2}{\epsilon'^2} \bigg(-\epsilon~\epsilon' (r^2 +4(\alpha+ \beta^2 + \beta Log[\frac{2 \beta \epsilon'}{r^2}]) \epsilon')+ 2(\alpha+ \beta+ \beta^2 + \beta Log[\frac{2 \beta \epsilon'}{r^2}]) \epsilon' (\epsilon- r \epsilon') (2 \epsilon'- r \epsilon'')- \nonumber\\
&& 4 (r - \epsilon) \bigg(2 (3 \alpha + \beta (5+ 3 \beta)+ 3 \beta Log[\frac{2 \beta \epsilon'}{r^2}]) \epsilon'^2 + r^2 \beta~ \epsilon''^2 + r~ \epsilon' (-4(\alpha + \beta (2+ \beta)+ \beta Log [\frac{2 \beta \epsilon'}{r^2}]) \epsilon'') + r (\alpha + \beta  \nonumber\\
&& + \beta^2 + \beta Log [\frac{2 \beta \epsilon'}{r^2}]~ \epsilon''' ) \bigg)        \bigg)- \bigg((\epsilon- r \epsilon') (r^2 + 4 (\alpha + \beta^2 + \beta Log [\frac{2 \beta \epsilon'}{r^2}])\epsilon' ) -8(\alpha + \beta +\beta^2 + \beta Log [\frac{2 \beta \epsilon'}{r^2}] )
\nonumber\\
&&  (r - \epsilon) (-2 \epsilon' + r \epsilon'' ) \bigg) \bigg],\label{ddd}
\end{eqnarray}
Now, we want to find shape function from this equation. It is a nonlinear differential equation and cannot be solved analytically. Thus, we find its numerical solution, shown in Fig. $\ref{Fig:55}$. It can be seen from Fig. $\ref{Fig:55}$ that our numerical shape function fulfills all the requirements and throat lies at $r_{0}=1.2$ Km as evident from the magnified view of second plot in Fig. $\ref{Fig:55}$. Subsequently, we observe that our wormhole solution respects WEC for throughout the wormhole geometry. It means the entire wormhole geometry is sustained on ordinary matter. Our observation shows that $df/dR >0 $ (first plot in Fig. $\ref{Fig:55}$), which means our solution represents non-asymptotically flat wormhole geometry in case of isotropic pressure with logarithmic correction.
As shown in third plot of Fig. $\ref{Fig:55}$, the rapid increase of pressure and density away from the throat and very large values of density $~10^{145} g/cm^3$ (which extremely exceed the Planck density) indicates that in this case the wormhole configuration will be at least unstable. In fact, it has been shown that in some situations no realistic wormholes can be formed, for example while studying viable scalar-tensor models of dark energy \cite{5bron}.
Moreover, some studies show that wormhole configurations may be unstable under non-static monopole (spherically symmetric) perturbations \cite{100,101,102}.

\subsection{Barotropic Equation of State}

\begin{figure}\center
\begin{tabular}{cccc}
\epsfig{file=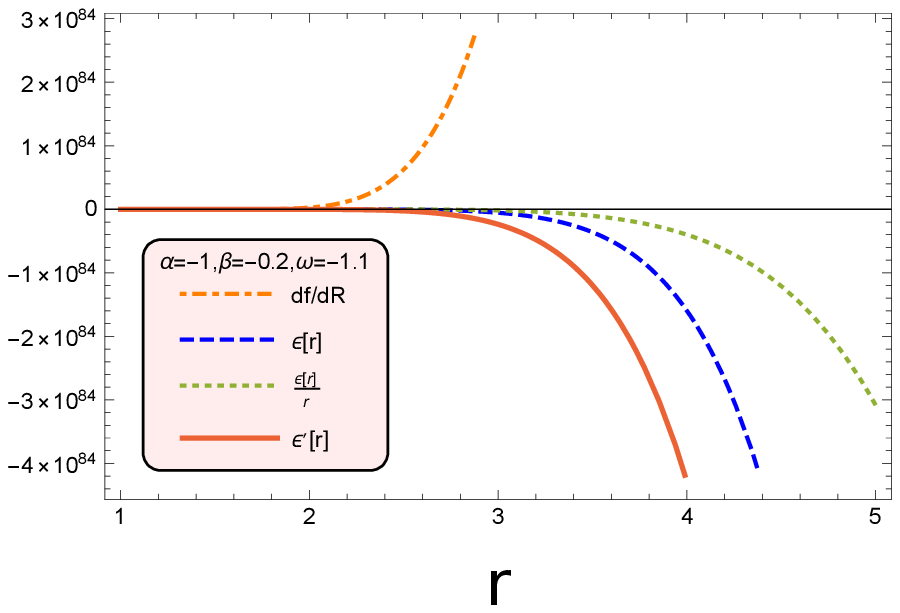,width=0.34\linewidth} &
\epsfig{file=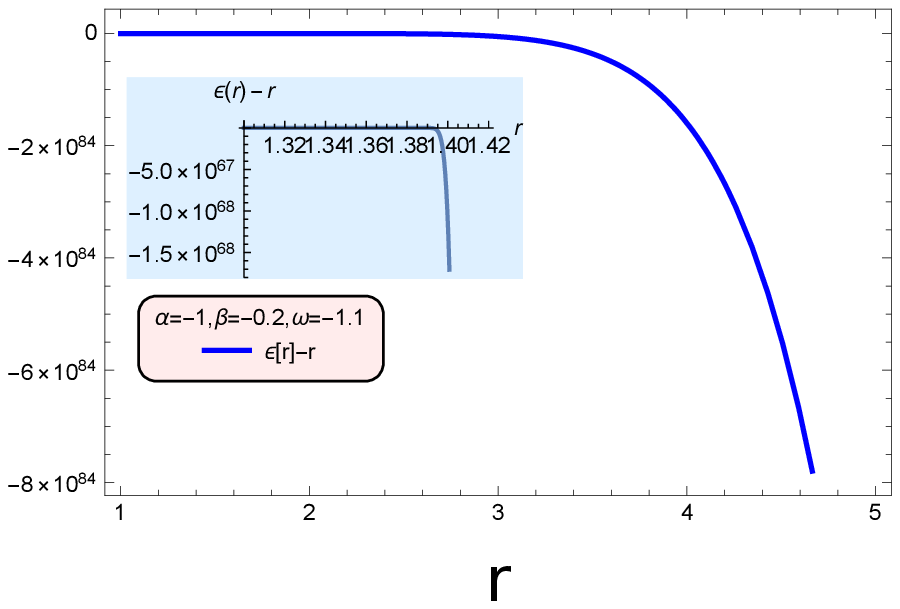,width=0.34\linewidth} &
\epsfig{file=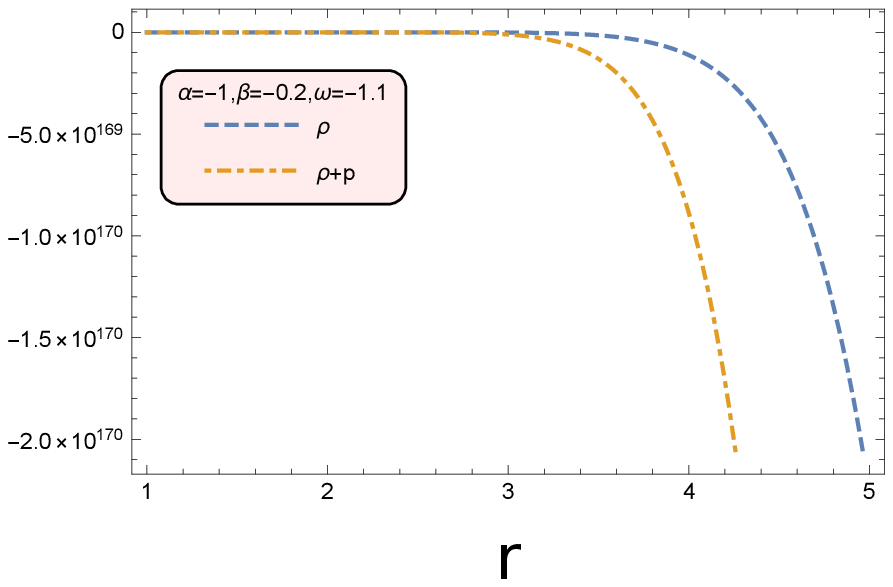,width=0.34\linewidth} \\
\end{tabular}
\caption{First plot shows the behaviour of $\epsilon-r$ (1st), $1-\frac{\epsilon}{r}$ and $\epsilon'$ by using $\alpha=-1,~~ \beta=-0.2$ and $\omega=-1.1$ for barotropic pressure. Second plot verify the violation of non-existence theorem and third plot shows that NEC is respected at the throat but WEC is violated at the throat.}\center
\label{Fig:57}
\end{figure}

Many researchers have investigated the wormhole solutions for barotropic fluid i.e, radial pressure is a function of density.
Here, we consider $p_r= \omega \rho$, where $ \omega $ is a constant. By employing this condition on Eqs. ($\ref{a}$) and ($\ref{b}$), we get

\begin{eqnarray}
&& \frac{1}{r^5 \epsilon'} \bigg[-\epsilon~\epsilon' (r^2 +4(\alpha+ \beta^2 + \beta Log[\frac{2 \beta \epsilon'}{r^2}]) \epsilon')+ 2(\alpha+ \beta+ \beta^2 + \beta Log[\frac{2 \beta \epsilon'}{r^2}]) \epsilon' (\epsilon- r \epsilon') (2 \epsilon'- r \epsilon'')-4 (r - \epsilon) \bigg(2 (3 \alpha +  \nonumber\\
&&  \beta (5+ 3 \beta)+ 3 \beta Log[\frac{2 \beta \epsilon'}{r^2}]) \epsilon'^2 + r^2 \beta~ \epsilon''^2 + r~ \epsilon' (-4(\alpha + \beta (2+ \beta)+ \beta Log [\frac{2 \beta \epsilon'}{r^2}]) \epsilon'') + r (\alpha + \beta + \beta^2 + \beta Log [\frac{2 \beta \epsilon'}{r^2}]~ \epsilon''' ) \bigg)        \bigg] \nonumber\\
&&
 = \frac{ \omega \epsilon'}{r^4} \bigg[r^2 +4(\alpha+ \beta^2 + \beta Log[\frac{2 \beta \epsilon'}{r^2}]) \epsilon'  \bigg].\label{aa}
\end{eqnarray}
This is again a nonlinear differential equation. So, in the same manner, we apply some numerical techniques and find its numerical solution.
Our presented shape function obeys all the physical requirements and we also verify the violation of non-existence theorem in neighborhood of throat which lies at $r_0 =1.39$ Km by using parameters $\alpha=- 1$, $\beta =-0.01$ (see in Fig. $\ref{Fig:57}$, plot $1$). The equation of state $p_r  / \rho= \omega$ for $\omega < -1$ dubbed as $``phantom~~ energy"$ and contains some strange properties \cite{5cal}. The presence of phantom energy causes the increase of density up to infinity in finite time \cite{5wang}. In the following work, we consider $\omega=-1.1$ and observe that energy density is negative at the throat and then goes to negative infinity outside the throat. The wormhole solution violates NEC and WEC in case of barotropic pressure (see last plot in Fig. $\ref{Fig:57}$) with logarithmic correction.

\section{Comparison and Conclusion}

In present paper, we examine suitable shape functions for the formation of relativistic wormholes in the frame work of logarithmic corrected $R^2$ gravity i.e.,
$f(R)=R+ \alpha R^2+ \beta R^2 Log(\beta R)$. To describe the relativistic wormholes geometry, it is necessary that shape function match all the physical
requirements (already discussed in section $1$). As for the existence of traversable wormhole, the red-shift parameter $\psi$ must be finite everywhere.
Moreover, the condition $\psi=d\rightarrow0$ as $r\rightarrow\infty$
suggests that $\psi$ should be asymptotically flat as well. To preserve these conditions, we fix the red-shift parameter as a constant function i.e, $\psi=d$,
throughout our work. For anisotropic case, different researchers use various types of shape functions to discuss static wormholes. Here, we examine wormholes
for anisotropic fluid with three different shape functions by assuming appropriate combinations of parameters $\alpha$ and $\beta$. The main results of present study are itemized below.\\\\
\begin{itemize}
\item Firstly, we consider shape function $\epsilon(r)=(r_0)^{n+1} r^{-n}$, where $n$ is an arbitrary constants. The violation of non-existence theorem for parameters $r_0=2$, $\alpha=20,~~ \beta =-0.1$ and $0 \leq n\leq 2$, have shown in Fig. $\ref{Fig:51}$. One can notice that $\frac{df}{dR}$ is negative at the throat which shows the existence of wormhole geometry. Further, we evaluate the energy conditions in Fig. $\ref{Fig:52}$. It can be seen that $\rho$ and $\rho+ p_r$  are negative at the throat and $\rho+ p_t$ is positive at the throat but it becomes negative outside the throat. We also observe that any combination of free parameters that violates the non-existence theorem, violates the WEC. Thus, we can say that one cannot avoid the presence of exotic matter for static spherically symmetric wormhole for shape function Eqn ($\ref{8}$) in the context of logarithmic corrected $f(R)$ model.
\item Secondly, we assume exponential shape function $\epsilon(r)= \frac{r}{e^{(r-r_0)}}$. One may notice that region plot in Fig. $\ref{Fig:53}$, for $r_0 =2$, $\alpha=20$ and $-2 \leq \beta \leq 0$, provides wormhole solution with the violation of non-existence theorem. Moreover, $\rho$ and $\rho +p_r$ are positive at the throat. Subsequently, we observed that wormhole geometry has positive energy density in space, whereas $\rho + p_t$ is also positive near the throat. Thus, we can say that for this particular choice of shape function, exotic matter can be avoided (or may be with less amount of exotic matter) for the construction of wormhole geometry.

\item Thirdly, we consider shape function  $\epsilon(r)= \frac{r_0 Log[1+r]}{Log[1+r_0]}$. It is clear from Fig. $\ref{Fig:566}$ that NEC is violated at the throat. Fig. $\ref{Fig:56}$ shows $df/dR <0$, which violates the condition for non-existence theorem. Further, we find that with the logarithmic corrections, one can not found the static spherically symmetric wormhole solutions without exotic matter which satisfies the non-existence theorem for any combination of free parameters for this particular shape function.

    Thus, one may conclude that for anisotropic case exponential shape function is more suitable as compare to other shape function. Exponential shape function provides traversable wormhole solutions with less amount of exotic matter in the context of logarithmic corrected $R^2$ gravity.

\item In the same manner, we can also find the wormhole solutions which are maintained principally by matter sources with isotropic pressure i.e., $p=p_{r}=p_{t}$. We have plot the numerical behaviour of shape function for isotropic case in Fig. $\ref{Fig:55}$ (first and 2nd plot) for $\alpha=0.00001$ and $\beta= 1$. In second plot of Fig. $\ref{Fig:55}$, we verify that $\frac{df}{dR} <0$ which provides static spherically symmetric solution in $f(R)$ gravity. Whereas, third plot in Fig. $\ref{Fig:55}$, shows that our solution respects WEC. Logarithmic correction provides the wormhole geometry filled with ordinary matter.

\item lastly, we also discuss wormhole solutions for barotropic fluid i.e, $\rho= \omega p_r= \omega p$. Our presented shape function obeys all the physical requirements (see in Fig. $\ref{Fig:57}$, plot $1$). We find that throat lies at $r_0 =1$ by using parameters $\alpha= -1$, $\beta =-0.2$. The equation of state $p_r  / \rho= \omega$ for $\omega < -1$ dubbed as $``phantom~~ energy"$ and contains some strange properties \cite{5cal}. The presence of phantom energy causes the increase of density up to infinity in finite time \cite{5wang}. In the following work, we consider $\omega=-1.01$ and observe that energy density is negative at the throat and then goes to positive infinity outside the throat. In Fig. ($\ref{Fig:57}$) (middle part), we also verify the violation of non-existence theorem.

    Thus, we can say that in barotropic case, exotic matter can not be avoided in the context of logarithmic corrected $R^2$ gravity.\\\\
\end{itemize}
Conclusively, logarithmic-corrected $R^2$ gravity model provides very interesting results which shows that this model can be helpful to find the solutions for many other cosmic issues like dark energy, dark matter halo, formation of black holes and strange stars. As black hole mergers produces gravitational waves may
further provide alluring constraints on parameters involved in modified gravity models.\\\\


\section*{References}
\begin{thebibliography}{70}

\bibitem{fla} Flamm, L.: Physikalische Zeitscrift \textbf{XVII}(1916) 448.\\

\bibitem{ein} Einstein, A. and Rosen, N.: Phys. Rev. \textbf{48} (1935) 73.\\

\bibitem{kim} Kim, S. W., Thorne, K. S.: Phys. Rev. \textbf{D43} (1991) 3929.\\

\bibitem{haw1} Hawking, S. W.: Phys. Rev. \textbf{D46} (1992) 603.\\

\bibitem{hoc1} Hochberg, D. and Visser, M.: Phys. Rev. \textbf{D56} (1997a) 4745.\\

\bibitem{hoc2} Hochberg, D. and Visser, M.: Phys. Rev. \textbf{D58} (1998b) 044021 .\\

\bibitem{spe} Spergel, D. N., et al.: Astrophys. J. Suppl. Ser. \textbf{148} (2003) 175.\\

\bibitem{per} Perlmutter, S., et al.: Astrophys. J. \textbf{483} (1997) 565.\\

\bibitem{qad} Qadir, A., Lee, H. W., and Kim, K. Y.: Int. J. Mod. Phys. \textbf{D26} (2017) 1741001.\\

\bibitem{cap1} Capozziello, S., Cardone, V. F., Carloni, S., and Troisi, A.: Int. J. Mod. Phys. \textbf{D12} (2003) 1969.\\

\bibitem{dim} Demianski, M., Piedipalumbo, E., Rubano, C., and Tortora, C.: Astron. Astrophys. \textbf{454} (2006) 55.\\

\bibitem{cap4} Capozziello, S.:  Int. J. Mod. Phys, \textbf{D11} (2002) 483.\\

\bibitem{buc} Buchdahl, H. A.: Mon. Not. R. Astron. Soc. \textbf{150}(1970)1.\\

\bibitem{mar} Martin, J., Ringeval, C., Vennin, V.: JCAP \textbf{2014.10} (2014) 038.\\

\bibitem{noj1} Nojiri, S., Odintsov, S. D.: Phys. Rept. \textbf{505} (2011) 59.\\

\bibitem{har1} Harko, T., Lobo, F. S., Nojiri, S. I. and Odintsov, S. D.: Phys. Rev. \textbf{D84(2)} (2011) 024020.\\

\bibitem{rah} Rahaman, F., Banerjee, A., Jamil, M., Yadav, A. K. and Idris, H.: Int J Theor Phys, \textbf{53} (2014) 1910.\\

\bibitem{har2} Harko, T., Lobo, F. S., Mak, M. K. and Sushkov, S. V.: Phys. Rev. \textbf{D87} (2013) 067504.\\

\bibitem{pav} Pavlovic, P., Sossich, M.: Eur. Phys. J. C  \textbf{75} (2015) 117.\\

\bibitem{bah} S. Bahamonde, M. Jamil, P. Pavlovic and M. Sossich.: Phys. Rev. \textbf{D94} (2016) 044041.\\

\bibitem{zub1} Zubair, M., Kousar, F. and Bahamonde, S.: Eur. Phys. J. Plus \textbf{133} (2018) 523.\\

\bibitem{5bron} Bronnikov, K. A., and Starobinsky, A. A.: JETP letters, \textbf{85(1)} (2007) 1.\\

\bibitem{bro} Bronnikov, K. A., Skvortsova, M. V., and Starobinsky, A. A.: Gravitation and Cosmology, \textbf{16(3)} (2010) 216.\\

\bibitem{mor} Morris, M. S. and Thorne, K. S: Am. J. Phys. 56 (1988) 395.\\

\bibitem{haw2} Hawking, S. W.  and Ellis, G. F. R.: (Vol. 1) Cambridge
University Press 1973 doi:10.1017/CBO9780511524646.\\


\bibitem{hoc4} Hochberg, D. and Visser, M: Phys. Rev. Lett. \textbf{81} (1998a) 746.\\

\bibitem{ray} Raychaudhuri, A: Phys. Rev. \textbf{106} (1957) 172.\\

\bibitem{star} Starobinsky, A. A.: Phys. Lett. \textbf{B91}(1980)99.\\

\bibitem{eliz} Elizalde, E., Odintsov, S. D., Oikonomou, V. K., Paul, T.: JCAP \textbf{2019} (2019) 017.\\



\bibitem{104} Odintsov, S. D., Oikonomou, V. K. and Sebastiani, L.: Nucl. Phys.  \textbf{B923} (2017) 608.\\

\bibitem{105} Liu, L. H., Prokopec, T. and Starobinsky, A. A.: Phys. Rev. \textbf{D98} (2018) 043505.\\

\bibitem{106} Kirillov, A. A., and Savelova, E. P.: Int. J. Mod. Phys. \textbf{D25} (2016) 1650075.\\

\bibitem{jam} Jamil, M., Momeni, D., Myrzakulov, R.: Eur. Phys. J. C \textbf{73} (2013) 2267.\\

\bibitem{shar2} Sharif, M. and Zahra, Z: Astrophys. Space Sci. \textbf{348(1)} (2013) 275.\\

\bibitem{sham3} Shamir, M. F. and Zia, S: Astrophys. Space Sci. \textbf{363} (2017) 247.\\

\bibitem{zub2} Zubair, M., Waheed, S., Ahmad, Y.: Eur. Phys. J. C  \textbf{76} (2016) 444.\\

\bibitem{lob} Lobo, F.S.N., Oliveira, M.A.: Phys. Rev. \textbf{D80} (2009) 104012.\\

\bibitem{sam} Samanta, G. C., Godani, N. and Bamba, K.: ( 2018).  arXiv:1811.06834.\\

\bibitem{5goda} Godani, N. and Samanta, G. C.: Int. J. Mod. Phys. \textbf{D28} (2018) 1950039.\\

\bibitem{5har1} Harko, T., Lobo, F. S., Mak, M. K., and Sushkov, S. V.: Phys. Rev. \textbf{D87(6)} (2013) 067504.\\

\bibitem{5li} Lin, R. H., Wu, Z. Y., and Zhai, X. H.: (2019). arXiv:1906.10323.\\


\bibitem{100} Starobinsky, A. A.: Sov. Astron. Lett. \textbf{7} (1981) 36.\\

\bibitem{101} Bronnikov, K. A. and Grinyok, S. V.:  Grav. Cosmol. \textbf{10} (2004) 237.\\

\bibitem{102} Bronnikov, K. A. and Grinyok, S. V.:  Grav. Cosmol. \textbf{11} (2005) 75.\\


\bibitem{5cal} Caldwell, R. R.: Phys. Lett. \textbf{B545} (2002) 23.\\

\bibitem{5wang} Wang, L., Caldwell, R. R., Ostriker, J. P. and Steinhardt, P. J.: Astrophys. J. \textbf{530} (2000) 17.\\

\end{thebibliography}
\end{document}